\documentclass[useAMS,usenatbib]{mn2e}
\usepackage{graphicx,amsmath,multirow,amssymb}
\usepackage{subfig}
\usepackage{natbib}
\newcommand{\comment}[1]{}

\def\simgt{\lower.5ex\hbox{$\; \buildrel > \over \sim \;$}}
\def\simlt{\lower.5ex\hbox{$\; \buildrel < \over \sim \;$}}

\title[AGB stars in the SMC]{AGB stars in the SMC: evolution and dust properties based on {\it Spitzer} observations}
\author[Dell'Agli et al.]{F. Dell'Agli$^{1,2}$, D. A. Garc\'{\i}a--Hern\'andez$^{3,4}$, P. Ventura$^2$, R. Schneider$^2$, 
\newauthor
M. Di Criscienzo$^{2}$, C. Rossi$^{1}$ \\
$^1$Dipartimento di Fisica, Universit\`a di Roma ``La Sapienza'', P.le Aldo Moro 5, 00143, Roma, Italy \\
$^2$INAF -- Osservatorio Astronomico di Roma, Via Frascati 33, 00040, Monte Porzio Catone (RM), Italy \\
$^{3}$Instituto de Astrof\'{\i}sica de Canarias, C/ Via Láctea s/n, E-38205 La Laguna, Tenerife, Spain \\
$^{4}$Departamento de Astrof\'{\i}sica, Universidad de La Laguna (ULL), E-38206 La Laguna, Tenerife, Spain\\
}

\begin{document}

\date{Accepted, Received; in original form }

\pagerange{\pageref{firstpage}--\pageref{lastpage}} \pubyear{2012}

\maketitle

\label{firstpage}

\begin{abstract}
We study the population of asymptotic giant branch (AGB) stars in the Small Magellanic
Cloud (SMC) by means of full evolutionary models of stars of mass 
$1~M_{\odot} \leq M \leq 8~M_{\odot}$, evolved through the thermally pulsing phase.
The models also account for dust production in the circumstellar envelope. We compare {\it Spitzer} infrared colours with results from theoretical modelling.

We show that $\sim 75\%$ of the AGB population of the SMC is composed by scarcely obscured
objects, mainly stars of mass $M \leq 2M_{\odot}$ at various metallicity, formed between 700 Myr and 5 Gyr ago; 
$\sim 70\%$ of these sources are oxygen--rich stars, while $\sim 30\%$ are C--stars.

The sample of the most obscured AGB stars, accounting for $\sim 25\%$ of the total sample, is composed almost entirely
by carbon stars. The distribution in the colour--colour ($[3.6]-[4.5]$, $[5.8]-[8.0]$)
and colour--magnitude ($[3.6]-[8.0]$, $[8.0]$) diagrams of these C--rich objects, with 
a large infrared emission, traces an obscuration sequence, according to the amount of carbonaceous dust in their 
surroundings. The overall population of C--rich AGB stars descends from $1.5-2~M_{\odot}$ stars of metallicity 
$Z=4\times 10^{-3}$, formed between 700 Myr and 2 Gyr ago, and from lower
metallicity objects, of mass below $1.5~M_{\odot}$, 2-5 Gyr old. 

We also identify obscured oxygen-rich stars ($M \sim 4-6M_{\odot}$) experiencing hot bottom burning.
The differences between the AGB populations of the  SMC and LMC are also commented.

\end{abstract}

\begin{keywords}
Stars: abundances -- Stars: AGB and post-AGB. ISM: abundances, dust 
\end{keywords}

\section{Introduction}
Stars evolving through the AGB phase are regarded as 
important dust manufacturers \citep{gehrz89}. The winds of AGB stars are an extremely
favourable environment to dust formation, owing to the high densities and the relatively
low temperatures, which favour condensation of gas molecules into dust grains 
\citep{gs85, gs99}. For these reasons, knowledge of dust production from AGB stars
proves important for a number of astrophysical contexts:
the interpretation of the spectral energy distribution (SED) 
of high--redshift quasars \citep{bertoldi03, wang08, wang13};
the study of dust evolution in galaxies of the Local Group \citep{dwek98, calura08,raffa14,matteo14};
the possible explanations of the presence of dust at early epochs \citep{valiante09,
valiante11, pipino11}. Furthermore,
the debate concerning the relative contributions from different kinds of stars to the
overall dust budget is still alive: the early claim of a dominant contribution from
SNe \citep{maiolino04} was challenged by following investigations, focused on the effects
of the reverse shocks on dust destruction \citep{bianchi07}. The important contribution
from AGB stars to dust production at high redshifts was outlined by \citet{valiante09}.

Modelling dust formation around AGB stars have made significant progresses in the
last few years. The pioneering investigations by the Heidelberg group 
\citep{fg01, fg02, fg06} were followed by works from other research teams
\citep{paperI, paperII, paperIII, paperIV, nanni13a, nanni13b, nanni14}. The results
found by the various groups present considerable differences in
the amount of dust produced by AGB stars and the kind of particles formed \citep{nanni13b}. 
This is due to the different input physics used, particularly for
what concerns the most relevant factors affecting AGB evolution, i.e. convection and
mass loss \citep{vd05a, vd05b, doherty14}.

Unfortunately, despite some admirable attempts \citep{canuto92, canuto93}, we are still far from an 
exhaustive description of the convective phenomenon, accounting for non locality, based 
on a self--consistent solution of the Navier--Stokes equations; this approach would be the 
only way to determine, based on first principles, the efficiency of the convective 
transport of energy and the mixing induced by convective eddies.

Therefore, the only possibility to achieve a more reliable description of the AGB phase, 
with a higher predictive power of the results obtained, is to fix the main properties of 
AGB stars via a detailed comparison with the observations. The Magellanic Clouds (MCs) are
the ideal environments to this aim, owing to the relatively short distance 
(51 Kpc and 61 Kpc, for the LMC and SMC, respectively, Cioni et al. 2000, 
Keller \& Wood 2006) and the low reddening ($E_{B-V}=0.15$~mag and 0.04 mag,
for the LMC and SMC, respectively, Wasterlund 1997). 
The study of MCs, AGB stars offers better observational constraints to the theoretical
models than their Galactic counterparts (e.g., Garc\'{\i}a--Hern\'andez et al. 2006, 2007, 2009) 
because of the unknown distances (and larger reddening) in our own Galaxy.
The {\it Spitzer Space Telescope} made available to the community data of millions of AGB stars
in the MCs: ``Surveying the Agents of a Galaxy Evolution Survey"
\citep[SAGE--LMC,][]{meixner06} and the Legacy program entitled ``Surveying the Agents of 
Galaxy Evolution in the tidally stripped, low metallicity Small Magellanic Cloud" 
\citep[SAGE--SMC,][]{gordon11} have provided spatially and photometrically complete
infrared surveys of the evolved star population in the MCs.

\citet{flavia15} (hereinafter D15), in a recent investigation, used dusty AGB models to 
study the AGB population of the LMC. The theoretical results were compared 
to {\it Spitzer} data from Riebel et al. (2012). D15 presented an interpretation of the observed 
sample of AGB stars, by characterizing the stars in terms of age, mass of the progenitors, 
chemical composition, dust in the circumstellar envelope. The comparison among the observed
and expected distribution of AGB stars in the colour--colour and colour--magnitude diagrams
obtained with the {\it Spitzer} infrared bands allowed the determination of the relative contributions
from C--stars and oxygen--rich AGB stars. The most relevant result was the study of the most
obscured sources, called `` extreme" : the stars were identified as a majority of C--rich
stars, in the final evolutionary AGB phases, and a smaller group of younger objects,
undergoing Hot Bottom Burning (hereinafter HBB). The study of D15 provided an important 
feedback on the details of AGB modelling, particularly for what concerns the depth of the
Third Dredge--Up (TDU). The possibility that the IR colours of massive AGB stars, associated with 
a measure of the C/O ratio, can be used 
to determine the efficiency of HBB, was investigated by \citet{ventura15}. 

In the present paper we extend to the SMC the analysis applied by D15 to the
LMC. The goal of the present investigation is to compare our interpretation of the SMC
population of AGB stars with the classification of the sources observed, present in the
literature. To this aim, similarly to D15, we attempt a characterization of the stars
observed to infer their mass, formation epoch, surface chemistry and the dust in their
surroundings. A particular attention is dedicated to the stars with the largest infrared
emission, to derive their contribution to dust production in the SMC. 
For this reason, we base the comparison among models and observations on the 4 IRAC filters
centered at $3.6\mu$m, $4.5\mu$m, $5.8\mu$m, $8.0\mu$m, covering the wavelength range
where most of the infrared emission from obscured, dusty AGB stars occurs.

The present investigation will be an important test for the theories regarding the evolution
of AGB stars and the dust formation process in their winds. The analysis by D15 showed that the
current generation of AGB models is able to satisfactorily reproduce the obscured AGB population in 
the LMC. To assess the reliability
of these models, it is crucial to understand whether the same description of convection and
mass loss used in D15, and the schematisation of the dust formation process adopted in that
work, allows the
description of the AGB population in a galaxy with a different total mass, star formation history, age-metallicity relations, as the SMC is.
Because the SFH of the LMC and SMC show important differences \citep{harris04,harris09}, the present work offers
the opportunity to test AGB models in a different range of mass and metallicity, with 
respect to those used by D15 \citep[see also][]{raffa14}.

This work is to be considered a first, important step, towards the description of the
AGB population in more distant environments, and to estimate their contribution to dust production. 
This study will help to be prepared in the future, to take advantage of the many, 
challenging opportunities, offered by the incoming observational facilities.

The paper is organized as follows: the data set of AGB stars in the SMC used for our analysis is
described in section 2; the numerical and physical input used to model the AGB 
phase and the dust formation process, and to produce the synthetic population of AGB stars are
given in section 3; section 4 presents the main results concerning the evolution
properties of AGB stars and of the dust in their surroundings; the interpretation of the 
observations is discussed in section 5, while a comparison of our results with the classification 
by \citet{boyer11} is given in section 6. Finally, Section 7 compares the
AGB populations in the SMC and LMC and our conclusions are
offered in Section 8.

\section{AGB stars in the SMC: observations and classifications}
\label{obs}
The SMC is a valuable environment to study evolved stars 
and their contribution to life cycle of dust in the Universe. This stems from the 
relative proximity  
($\sim 61 kpc$) and the low ISM metallicity.
Several SMC infrared surveys have been conducted so 
far: IRAS (Schwering \& Israel 1989; Miville-Deschenes \& Lagache 2005), the 
{\it Infrared Space Observatory (ISO)} (Wilke et al. 2003), and MSX (Price et al. 2001). 
The growing interest towards this galaxy has stimulated more recent near-IR and mid-IR 
surveys of the SMC: the AKARI survey (Ita et al. 2010) of 
small selected regions within the SMC bar and the {\it Spitzer} Survey of the Small Magellanic 
Cloud (S3MC; Bolatto et al. 2007), which imaged the SMC bar.

The photometric data from the {\it Spitzer} Legacy program ``Surveying the Agents of Galaxy 
Evolution in the SMC" \citep[SAGE-SMC][]{gordon11} provides a  
high--resolution, uniform, unbiased survey of the whole galaxy, including bar, wing 
and tail. Images were obtained in a 30 deg$^2$ field with IRAC (3.6, 4.5, 5.8, and 
8 $\mu$m) and MIPS (24, 70, and 160 $\mu$m) at two epochs. 

\citet{boyer11} used the SAGE-SMC survey to investigate the infrared properties of the 
evolved stars in the SMC. On the basis of the position in photometric planes, they classified 
AGB stars into four groups: oxygen--rich stars (O-AGB), anomalous oxygen--rich 
(aO-AGB), carbon stars (C-AGB) and ``extreme" stars (X-AGB). C-AGB and O-AGB stars were 
classified in the $J - K_s$ vs $K_s$ plane, following the cuts proposed by  
\citet{cioni06} and rescaling for metallicity and distance, as in \citet{cioni06b} 
\citep[see also section 3.1.1 in][]{boyer11}. To exclude contamination from RGB stars, 
the authors considered only objects brighter than the tip of the red giant branch 
(TRGB), with $K<K_s^{TRGB}=12.58$ mag \citep{cioni00} and $[3.6]<[3.6]_{TRGB}=12.6$ mag.  
The X-AGB are the most obscured AGB stars in the sample. Owing to their efficiency in 
producing dust and their infrared emission, part of these objects fall below the 
cuts used for the C-AGB and O-AGB \citep[see section 3.1.2 in][]{boyer11}. Therefore, 
these sources were defined as those brighter than $[3.6]_{TRGB}$ and with  $J - [3.6] > 3.1$. 
If the near--IR radiation is completely obscured by dust, stars with $[3.6]-[8.0] > 0.8$ are 
included in the X-AGB. Equation (1) and (2) in \citet{boyer11} describe the cuts 
used to minimize the contamination from young stellar objects (YSOs) and unresolved background galaxies.

In this paper, we compare the results from our theoretical models with the AGB sample presented 
in \citet{boyer11}, following the selection described above. 
Our analysis is focused on the IRAC filters, therefore we exclude from the observational sample by \citet{boyer11} the objects which are missing the detection in at least one of these bands, which account for 1$\%$ of the total.
The resulting sample is constituted by $\sim 5700$ stars.

We further tested our interpretation of the AGB population of the SMC against spectroscopically confirmed samples, observed in the last decades. Our analysis is mainly based on the recent work by \citet{ruffle15}, who presented a detailed classification of 58 AGB stars of the SMC, using spectra taken with the {\it Spitzer Infrared Spectrograph}. 
We also compare our predictions with the stars analyzed by \citet{smith95} and the results by \citet{anibal09}, that presented high-resolution optical spectra (lithium and/or s-process element
abundances) of a sample of unobscured and obscured AGB stars in the Magellanic
Clouds. For these samples we did a cross--correlation with the observations by \citet{boyer11}, to obtain the {\it Spitzer} magnitudes.

\section{The simulation of the AGB population of the SMC}
\label{inputs}
To compare the expected and observed distribution of stars in the various 
colour--colour and colour--magnitude diagrams obtained with the different IRAC filters, 
we followed a synthetic approach.

\begin{figure}
\resizebox{1.\hsize}{!}{\includegraphics{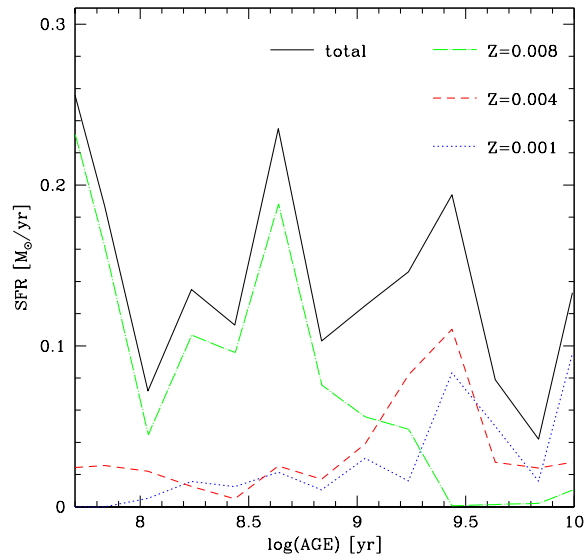}}
\vskip-50pt
\caption{The star formation history of the SMC as a function of stellar age (present time is $t=0$), according to
\citet{harris04}. The solid (black) line gives the total SFH, whereas the
dotted (blue), dashed (red), dotted--dashed (green) tracks give the contributions
from stellar populations of metallicity, respectively, $Z=10^{-3}$, $Z=4\times 10^{-3}$ 
and $Z=8\times 10^{-3}$.
}
\label{fsfh}
\end{figure}

We assume the Star Formation History (SFH) of the SMC given by \citet{harris04}. A 
plot of the variation with time of the Star Formation Rate (SFR) is given in Fig. \ref{fsfh}. Note that
in the work by \citet{harris04} the total SFH is split into 3 stellar components, of
metallicity $Z=10^{-3}$, $Z=4\times 10^{-3}$
and $Z=8\times 10^{-3}$.

\subsection{Stellar evolution modelling}
\label{agbmodel}
The evolutionary sequences of central stars were calculated by means of the ATON code for 
stellar evolution \citep{mazzitelli89}. The numerical structure of the code is described 
in details in \citet{ventura98}, whereas the latest updates are given in \citet{ventura09}.

As stated previously, we used three sets of models, with metallicity
$Z=10^{-3}$, $Z=4\times 10^{-3}$ and $Z=8\times 10^{-3}$. The initial helium was 
$Y=0.25$ for the most metal--poor population, and $Y=0.26$ for $Z=4\times 10^{-3}$ and $Z= 8\times 10^{-3}$. 
The relative percentages of
the various chemical species are taken from \citet{gs98}. We assumed an alpha--enhancement
$[\alpha / Fe]=+0.4$ for the $Z=10^{-3}$ case, and $[\alpha / Fe]=+0.2$ for the higher
metallicities. 

All the evolutionary sequences were followed from the pre--main sequence, until the
almost total consumption of the external envelope.

We recall here the main physical input used in the computations, most relevant for the
present work.

The convective instability was described according to the Full Spectrum of Turbulence
(hereinafter FST) model, discussed in details in \citet{cm91}. Convection modelling is the most relevant
factor in determining the time scale of the AGB evolution and the modification of the
surface chemistry. As shown by \citet{vd05a}, use of the FST description makes massive
AGB stars, with initial mass above $\sim 3 M_{\odot}$, to experience strong HBB. 
This has the double effect of shortening the evolutionary times and to modify
the surface chemistry, according to the equilibria of proton--capture nucleosynthesis.  

Mixing in convective regions is treated as a non--instantaneous process, coupled with 
nuclear burning. For each chemical species we solve a diffusion--like equation,
following the schematization by \citet{cloutmann}. This approach demands the computation of convective
velocities, entering the diffusive coefficient. Within this framework, the overshoot
phenomenon, i.e. the penetration of convective eddies in regions formally stable against
convective motions, is described via the decay of velocities beyond the formal border,
fixed via the Schwarzschild criterion. We use an exponential decay of velocities into
radiatively stable regions, with an e--folding distance $l=\zeta H_p$, where the pressure
scale--height, $H_p$, is calculated at the formal border of convection. During the core
hydrogen and helium burning phases, we used $\zeta=0.02$ to describe overshoot from the
border of the convective core and the base of the stellar envelope; this is in agreement
with the calibration of the width of the main sequences of open clusters, given in
\citet{ventura98}. During the thermal pulses phase, we used $\zeta=0.002$ to model
extra--mixing from the bottom of the surface mantle and the borders of the convective
shell which forms in conjunction with the ignition of each thermal pulse; the latter value
is based on the calibration of the luminosity function of carbon stars in the LMC
\citep{paperIV}.

The description of mass loss is also a crucial ingredient for the modelling of the AGB
phase. The rate of mass loss determines the duration of the whole AGB evolution, the
number of thermal pulses experienced and, consequently, the modification of the 
surface chemistry \citep{vd05b, doherty14}. For oxygen--rich stars we
used the description of mass loss by \citet{blocker95}, based on hydrodynamic 
simulations by \citet{bowen88}. For carbon stars, we used the formulae giving the mass 
loss rate as a function of luminosity and effective temperature published by the Berlin 
group \citep{wachter02, wachter08}. This treatment is based on pulsating hydrodynamical 
models, in which mass loss is driven by radiation pressure on dust grains.

For what concerns the modelling of carbon stars, the computation of the low--temperature
molecular opacities for carbon--rich mixtures has a great impact on the results.
This is because when the C/O ratio becomes larger than unity, the formation of CN molecules
favours a considerable increase in the molecular opacities of the external layers of
the star; this, in turn, leads to an overall expansion of the whole outer region of the
stars, and to an increase in the rate of mass loss. Under these conditions, the star looses
the external mantle very rapidly, thus limiting the number of thermal pulses experienced
\citep{marigo02}. The interested reader may find in \citet{vm09, vm10} a detailed discussion
on this argument, and the relative implications for the AGB evolution.

In the models used here we calculated the molecular opacities in the stellar surface 
layers (temperatures below 10,000K) by means of the AESOPUS tool \citep{marigo09}.
The tables generated with the AESOPUS code are available in the range of
temperatures $3.2 \leq \log T \leq 4.5$. The reference tables assume the same initial 
composition of the models used in the present work. For each combination of 
metallicity and $\alpha-$enhancement, additional tables are generated, in which the reference
mixture is altered by varying the abundances of C, N and O. This step is done by introducing 
the independent variables $f_C$, $f_N$, $f_{CO}$, that correspond to the enhancement
(in comparison with the initial stellar chemistry) of carbon, nitrogen and of the 
C/O ratio, respectively.

\subsection{The formation and growth of dust grains}
\label{dustmodel}
The winds of AGB stars are a favourable environment for the formation and growth of
dust grains. This stems for the low temperatures of their surface layers, which
allow the formation of dust grains close to the stellar surface, at typical distances of 
$\sim 1-10R_*$ ($R_*$ is the stellar radius): in those regions the gas densities are 
sufficiently large to allow dust formation in great quantities \citep{gs85, gs99}.

The model for the growth of dust grains in the circumstellar envelopes of AGB stars
is described in details in the previous papers by our group \citep{paperI, paperII,
paperIII, paperIV}, where all the relevant equations, describing the thermodynamic
structure of the wind and the rate of growth of the various dust species, are given.

The outflow is spherically symmetric, and it expands isotropically from the surface of 
the star, with an initial velocity $v_0=1 km/s$. The formation of dust particles
favours the acceleration of the wind, owing to the action of radiation pressure on the 
dust grains. The latter effect depends on the extinction coefficient, describing the
interaction between the radiation from the star and the solid particles of a given 
species.

\begin{table*}
\begin{center}
\caption{The dust species considered in the present work} 
\begin{tabular}{l|l|l|c|l|}
\hline
Dust species & Formula & environment & key--species & optical constants   \\ 
\hline
Olivine          & $Mg_2SiO_4$  &  O--rich            &  Si  &   \citet{ossenkopf92}  \\ 
Pyroxene         & $MgSiO_3$    &  O--rich            &  Si  &   \citet{ossenkopf92}  \\ 
Quartz           & $SiO_2$      &  O--rich            &  Si  &   \citet{ossenkopf92}  \\ 
Corundum         & $Al_2O_3$    &  O--rich            &  Al  &   \citet{koike95}      \\ 
Iron             & Fe           &  O--rich; C--rich  &  Fe  &   \citet{ordal88}      \\
Carbon           & C            &  C--rich           &  C   &   \citet{hanner88}     \\ 
Silicon carbide  & SiC          &  C--rich           &  Si  &   \citet{pegourie88}   \\ 
\hline
\end{tabular}
\end{center}
\label{tabdust}
\end{table*}

The kind of particles formed is determined by the surface C/O ratio. 
In the winds of oxygen--rich stars formation of silicates occurs, at a distance that, 
depending on the effective temperature, is in the range $d \sim 5-10R_*$ from the stellar 
surface; also small quantities of solid iron are present. The most stable species is alumina dust ($Al_2O_3$),
which forms at a typical distance of $\sim 1-3 $ stellar radii from the stellar surface \citep{flavia14b}. In carbon--rich 
environments we find a similar situation, with a stable and extremely 
transparent dust species, silicon carbide (SiC), forming close to the surface of the 
star ($d \sim 1-2R_*$), surrounded by a more external region ($d \sim 5R_*$), where solid 
carbon grains form and grow.

The amount of a given dust species which can be formed depends on the surface mass 
fraction of the so called ``key--element", i.e. the least abundant chemical species, 
concurring in the condensation process.

A list of the various dust species used in the present work, with the list of key--species
and the references for the optical constants adopted, are given in Table 1.

The description of the wind is interfaced with the AGB evolution modelling, because 
the results obtained depend on the main physical quantities of the central stars, namely
effective temperature, mass loss rate, luminosity, surface gravity.

\subsection{Synthetic spectra}
\label{spectramodel}
The evolutionary models of AGB stars provide the variation of the main physical and
chemical quantities of the star during the whole thermal pulses phase, allowing
to follow the behaviour of luminosity, effective temperature, rate of mass loss, 
and of the surface chemical composition. The application of the dust formation model,
described in section \ref{dustmodel},
leads to the determination of the amount of dust formed in the wind, distributed
among the different dust species, reported in Table 1. The knowledge of the density
and grain size stratification of the wind allows the computation of the optical depth,
which gives an indication of the degree of obscuration of the radiation emitted from
the surface of the star.

All these ingredients are used to derive the spectral energy distribution of the star
for some selected models along the evolutionary sequence. To this aim, we used the code
DUSTY \citep{dusty}. 

The input radiation from the central star was obtained by interpolating
in gravity and effective temperature among the appropriate tables of the same metallicity:
we used the NEXTGEN atmospheres \citep{nextgen} for oxygen--rich stars, whereas
the COMARCS atmospheres \citep{grams} were adopted for carbon stars; in the latter case
we interpolated among the C/O values.

The effects of dust on the redistribution of the radiation from the star was calculated
by accounting for the effects of two dusty layers. In the innermost regions we consider
the contribution from the most stable dust species, i.e alumina dust for oxygen--rich
AGB stars and SiC for carbon stars. The radiation emerging from this more internal zone is
further reprocessed by a more external dusty layer, populated by grains of alumina dust
and silicates in case of stars with a surface $C/O<1$; for carbon stars, this more
external region is populated by solid particles of SiC and amorphous carbon.

Convolution of the emerging flux with the transmission curves of the different bands
allows calculating the various magnitudes and colours.

\section{The evolution of AGB stars}
The present work is based on the models of AGB evolution and of the dust formation process
in the wind, used in D15. Section 3 of D15 provides an exhaustive description of the evolution
of AGB stars of different mass, in the range of metallicities of interest here, and of the change in their spectral
energy distribution during the various AGB phases: this allows the description of the path
traced by the evolutionary tracks in the colour--colour and colour--magnitude planes.
An overview of the main results discussed in D15 is given in the following.

\subsection{AGB stars: physical properties and the surface chemistry}
The main evolution properties of the models used here are discussed in details in 
\citet{ventura13} for the metallicities $Z=10^{-3}$ and $Z=8\times 10^{-3}$, and in 
\citet{ventura14} for $Z=4\times 10^{-3}$. The interested reader is referred to these 
two papers for a detailed description of the AGB phase. Here we give a 
summary of the main evolutionary properties.

During the AGB phase the stars undergo a series of thermal pulses, when ignition of
$3\alpha$ reactions in a helium--rich buffer above the degenerate core occurs, and the 
CNO burning shell is temporarily extinguished. The external envelope is gradually
lost by stellar wind. Eventually, nuclear reactions in the H--burning shell are extinguished,
and the evolutionary tracks first moves to the blue region of the
Hertzsprung-Russell diagram, before
beginning the White Dwarf cooling. This event marks the end of the AGB phase, thus
determining its overall duration.

The luminosity of AGB stars increases with the initial mass, $M_{init}$, of the star: 
the higher is $M_{init}$, the higher is the mass of the degenerate CO core\footnote{Indeed, 
models with mass in the range $6.5M_{\odot} \leq M \leq 8M_{\odot}$ develop a core made up 
of oxygen and neon, after an off--center carbon ignition, and the following development 
of a convective flame, propagating inwards, until reaching the center of the star.}, the 
brighter is the star \citep[see left panel of Fig.~1 in][]{ventura13}. The luminosities of
AGB stars cover a range of approximately one order of magnitude, from $L \sim 10^4 L_{\odot}$
($M_{init} \sim 1~M_{\odot}$), to $L \sim 10^5 L_{\odot}$, for 
$M_{init} \sim 7.5-8~M_{\odot}$

This reflects into a difference in the duration of the whole AGB phase, depending on the
stellar mass \citep[see Table 1 in][]{flavia15}. Our computations indicate that 
the AGB phase of low--mass AGB stars lasts $\sim 1$Myr, whereas their counterparts of highest mass 
loose the external envelope within a few tens of kyr. We reiterate here that the evolution times are rather 
uncertain, as they depend on the input physics used to calculate the evolutionary 
sequences \citep{vd05a, doherty14}.

The surface chemical composition changes during the AGB life, according to the relative
contributions of TDU and HBB. The former favours a gradual increase 
in the surface carbon, that eventually can lead to the formation of a carbon star.
HBB consists in the ignition of nuclear burning in the innermost regions of the
convective envelope; the resulting change in the surface chemistry reflects the
equilibria of proton--capture nucleosynthesis.
HBB is activated when the temperature at the bottom of the convective envelope exceeds
$\sim 40MK$. This demands a minimum core mass of $\sim 0.8~M_{\odot}$, which
reflects into a constrain on the initial mass of the star: $M_{init} > 3~M_{\odot}$
\citep[see right panel of Fig.~1 in][]{ventura13}. The most prominent effect of HBB is
the destruction of the surface carbon, via proton capture.

This introduces a dichotomy in the evolution of the surface chemistry of AGB stars:

\begin{itemize}
\item{
Low--mass AGB stars, $1~M_{\odot} < M_{init} < 3~M_{\odot}$\footnote{The threshold mass separating
stars evolving as carbon stars from more massive objects, experiencing HBB, changes with
the metallicity. It is $3~M_{\odot}$ for $Z \geq 4\times 10^{-3}$, whereas it is
$\sim 2.5~M_{\odot}$ for lower metallicity stars \citep[see, e.g,][]{ventura13}}, 
evolve as carbon stars. The enrichment in
carbon depends on the number of TDU episodes experienced, and is therefore higher
for larger values of $M_{init}$: models with $M_{init} \sim 2.5-3~M_{\odot}$ achieve the
largest surface abundances of carbon. The ejecta of these stars will therefore be 
enriched in carbon and, at a lower extent, in oxygen.
}

\item{
High--mass AGB stars, with $M_{init} > 3M_{\odot}$, experience HBB, thus remaining oxygen--rich. The temperatures 
of the regions close to the bottom of the convective envelope increase with $M_{init}$ 
\citep[see right panel of Fig.~1 in][]{ventura13}. However, the trend of the degree of 
the nucleosynthesis experienced with $M_{init}$ is not straightforward, as it depends
on the delicate interplay between the temperatures in the envelope and the rate of
mass loss \citep{ventura11}. The gas expelled by massive AGB stars will be greatly enriched
in nitrogen, and reduced in carbon and oxygen \citep[see Figure 1 in][]{ventura14}. 
}

\end{itemize}

The modification of the surface chemistry is sensitive to the metallicity of the star. 
Models of smaller
metallicity experience stronger HBB, thus the proton--capture nucleosynthesis occurs
at higher temperatures, provoking a greater change in the mass fractions of the various
elements involved. This not only reflects into the CNO elements, as shown in 
Figure 1 in \citet{ventura14}, but also into the production of sodium and aluminium 
\citep[see Fig. 2 and 3 in][]{ventura14}. Turning to the AGB stars of lower mass,
the C--star stage is reached more easily in models of smaller metallicity, owing to the
lower initial content of oxygen in the gas from which the star forms.

\begin{figure*}
\begin{minipage}{0.33\textwidth}
\resizebox{1.\hsize}{!}{\includegraphics{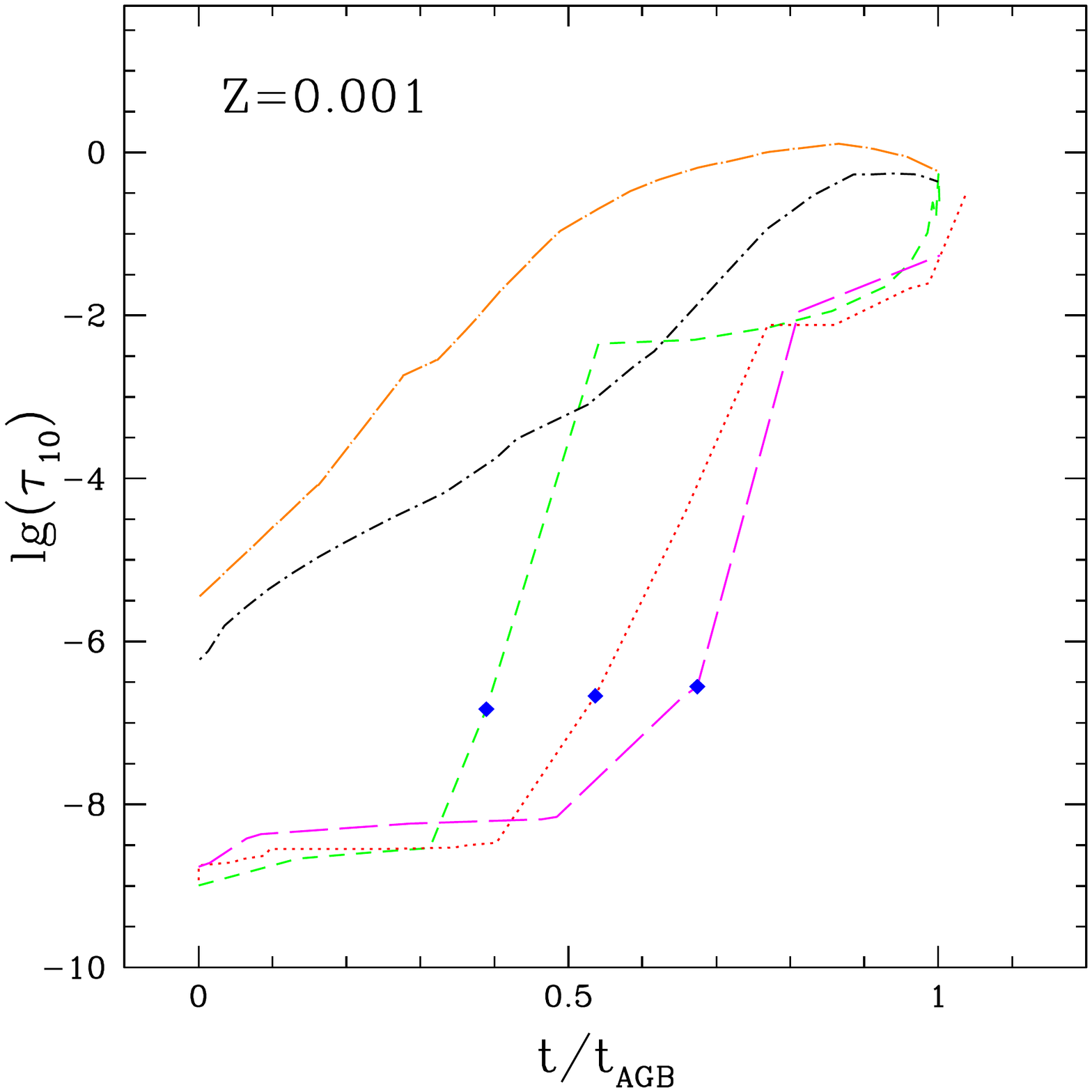}}
\end{minipage}
\begin{minipage}{0.33\textwidth}
\resizebox{1.\hsize}{!}{\includegraphics{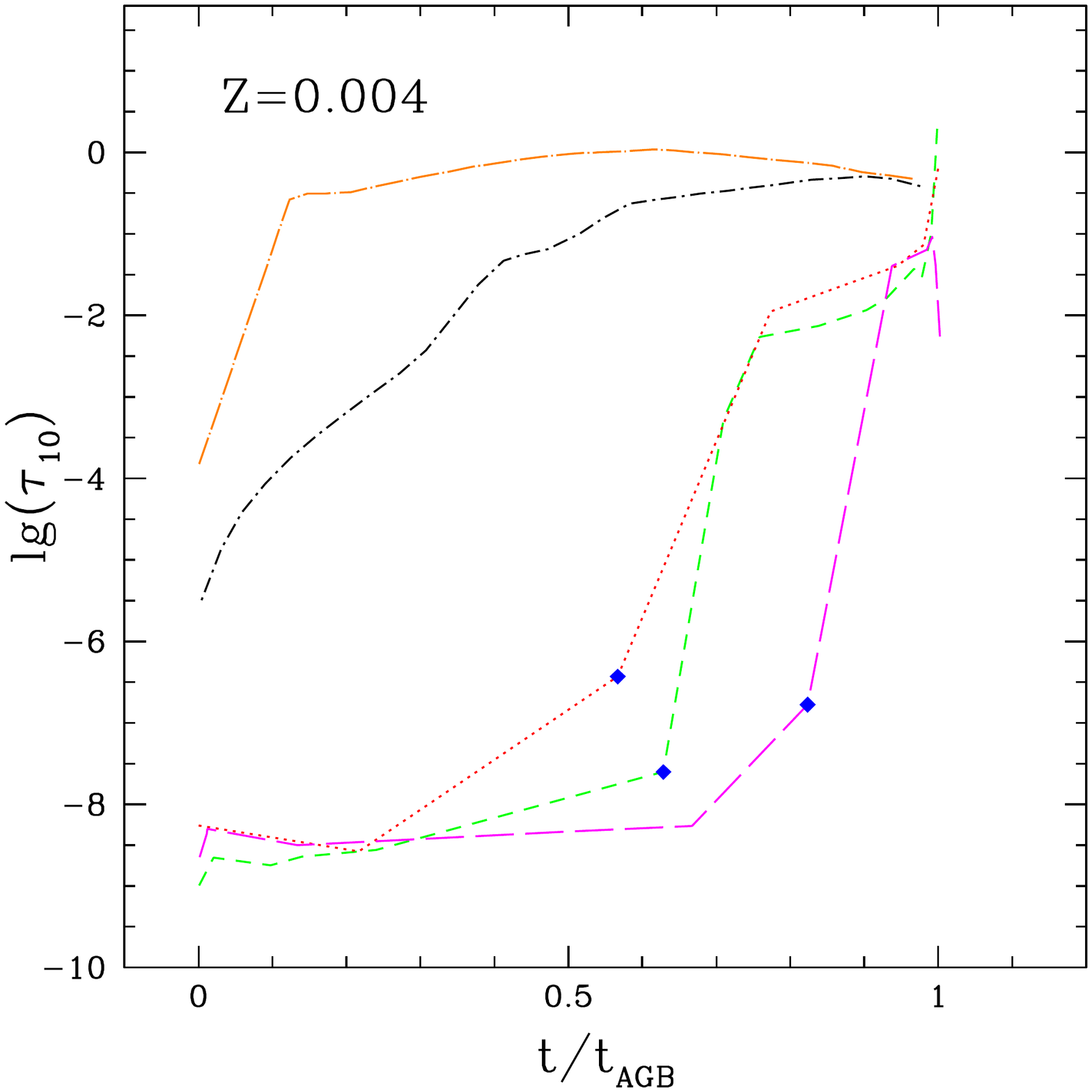}}
\end{minipage}
\begin{minipage}{0.33\textwidth}
\resizebox{1.\hsize}{!}{\includegraphics{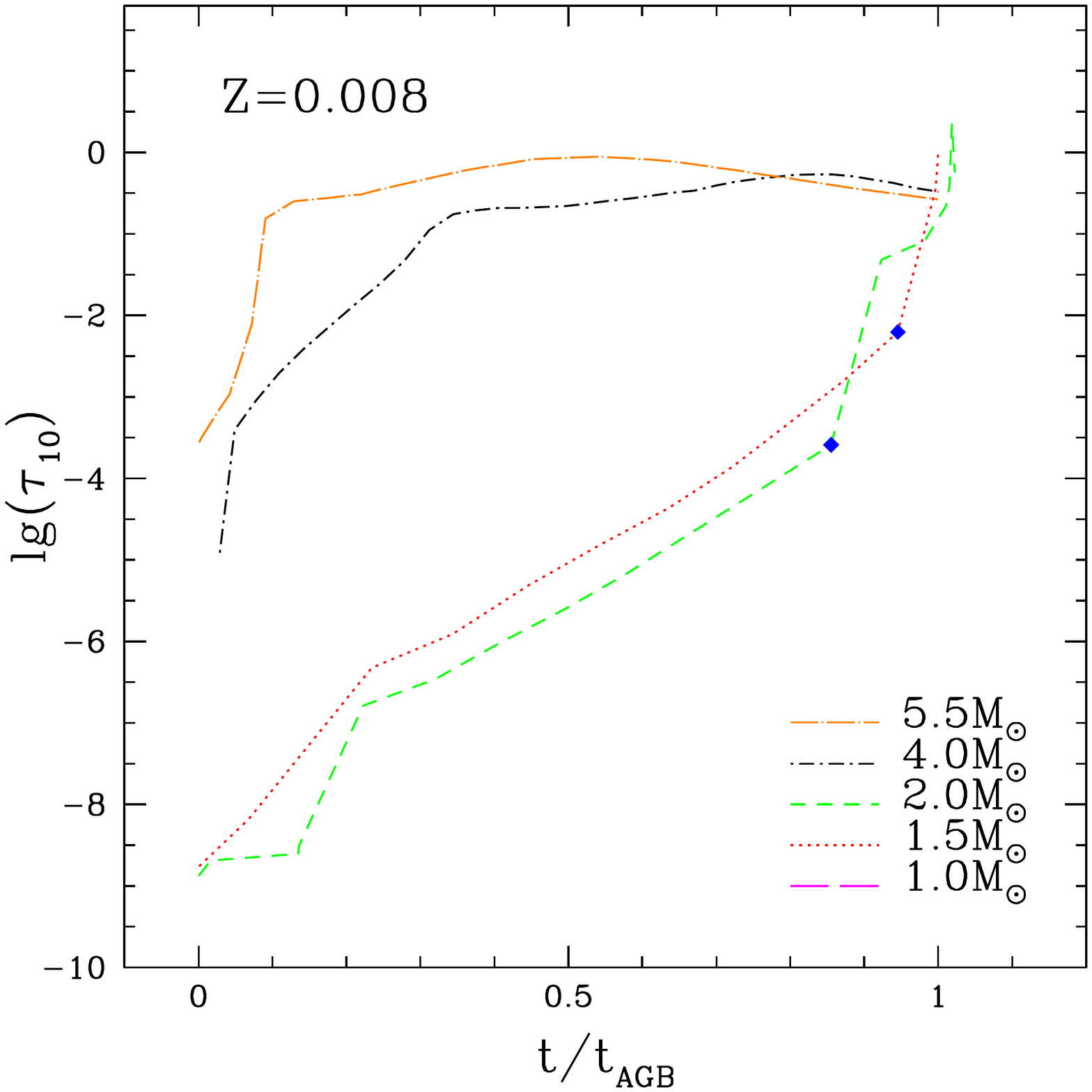}}
\end{minipage}
\vskip-30pt
\caption{The variation during the AGB evolution of the optical depth, $\tau_{10}$,
of AGB models of different initial mass and metallicity $Z=10^{-3}$ (left panel),
$Z=4\times 10^{-3}$ (middle) and $Z=8\times 10^{-3}$ (right). Times on the abscissa are
normalised to the total duration of the AGB phase. Blue squares along the tracks mark
the beginning of the C--star phase. In the right panel we do not show $1 M_{\odot}$ model at $Z=8\times 10^{-3}$, which does not become a C--star.
}
\label{ftau}
\end{figure*}

\subsection{Dust from AGB stars}
\label{dustagb}
As discussed in section \ref{dustmodel}, the latest generation of AGB models involve also
dust formation in the circumstellar envelope. The models published so far rely on the
schematisation by the Heidelberg group \citep{fg06}, where the growth of dust grains
takes place in an isotropically, expanding wind, moving outwards from the surface of the 
star. This schematisation was used in the recent explorations by \citet{nanni13a, nanni13b,
nanni14} and in the previous works on this argument by our group \citep{paperI, paperII,
paperIII, flavia14b}.

The dependence of dust production on the initial mass and the metallicity of AGB stars was
described in details in \citet{paperIV}, where the uncertainties affecting the results are also discussed. 

Based on the arguments of the previous section, SiC and
solid carbon grains form in the winds of
AGB stars of initial mass $1~M_{\odot} < M_{init} < 3~M_{\odot}$, whereas AGB stars of higher mass will be surrounded by
silicates and alumina dust grains.

The amount of SiC formed in low--mass AGB stars scales with metallicity, as the 
corresponding key--element, silicon, is proportional to Z. We find that the
typical size of SiC grains formed is $0.05\mu$m, $0.07\mu$m, $0.08\mu$m, for
AGB stars of metallicity $Z=10^{-3}$, $Z=4\times 10^{-3}$, $Z=8\times 10^{-3}$
\citep[see Fig. 5 in][]{paperIV}. The mass of silicon carbide produced is
$M_{SiC} \sim 10^{-5}-4\times10^{-5} M_{\odot}$ for $Z=10^{-3}$, 
$M_{SiC} \sim 3\times 10^{-5}-3\times 10^{-4} M_{\odot}$ for $Z=4\times 10^{-3}$,
$M_{SiC} \sim 10^{-4}-10^{-3} M_{\odot}$ for $Z=8\times 10^{-3}$
\citep[Fig. 4 in][]{paperIV}.

In C--rich environments the dust species produced in greatest quantities is solid carbon, 
despite being less stable than SiC: the reason is the much larger availability of carbon 
compared to silicon in the atmosphere of carbon stars. The amount of carbon--dust formed 
increases with the stellar mass, because stars of higher mass experience more TDU 
episodes, thus achieve a greater carbon enrichment at the surface. Stars of initial mass 
$M_{init} \sim 2-2.5~M_{\odot}$ are therefore the most efficient producers of carbon--dust, 
with grain sizes of the order of $\sim 0.2\mu$m in the circumstellar envelope 
\citep[see Fig. 5 in][]{paperIV}. This results is approximately independent of 
metallicity, because the carbon accumulated at the surface is synthesised in the 
He--burning shell, and is independent of Z. The overall mass of carbon dust produced by low--mass
AGB stars ranges from $M_d \sim 10^{-3} M_{\odot}$ for stars of initial mass 
$M_{init} \sim 1~M_{\odot}$, to $M_d \sim 10^{-2} M_{\odot}$ for  
$M_{init} \sim 2-2.5~M_{\odot}$ \citep[see Fig. 3 in][]{paperIV}.

In oxygen--rich stars, the mass of dust formed depends on the metallicity, because the
key--elements of the dust species formed, alumina dust and silicates, are aluminium 
and silicon, both dependent on Z. The size of the alumina grains formed are in the
range $0.03-0.07 \mu$m, according to $M_{init}$ and Z. The mass of alumina dust produced
is below $\sim 10^{-4} M_{\odot}$ for $Z=10^{-3}$, whereas it is 
$10^{-5} M_{\odot} < M_{Al_2O_3} < 10^{-3} M_{\odot}$ for the higher metallicities
\citep[see right panel of Fig. 9 in][]{paperIV}.
Most of the dust produced by oxygen--rich AGB stars is under the form of silicates, because
the silicon content largely exceeds aluminium in the surface layers. The size of the
dust grains formed increases with mass and metallicity. The silicates with the largest
size ($\sim 0.15\mu$m) form in the winds of massive AGB stars of metallicity $Z=8\times 10^{-3}$
\citep[Fig. 8 in][]{paperIV}. The overall mass of silicates produced is 
$10^{-4}M_{\odot}-10^{-3}M_{\odot}$ for $Z=10^{-3}$,
and in the range $10^{-3}M_{\odot}-10^{-2}M_{\odot}$ for $Z=4,8 \times 10^{-3}$.

\subsection{The infrared spectra of AGB stars}
\label{irspectra}
Understanding the dust formation process in the wind of AGB stars is crucial to interpret
the spectra of these stars, because the radiation emitted from the central star is
reprocessed by dust particles in the infrared. The shape of the spectral energy
distribution is determined by the dust species in the circumstellar envelope and by
the optical depth, indicating the degree of obscuration of the star. Here we use 
$\tau_{10}$, the optical depth at the wavelength of $10\mu$m.

Fig. \ref{ftau} shows the evolution of $\tau_{10}$ for the models used in this work;
each panel corresponds to a single metallicity. For clarity reasons, we show only
some of the masses involved in the present analysis. To show all the tracks in the same
plot we use as abscissa the evolution time, normalised to the overall duration of the
AGB phase.

Low-mass stars evolve initially as oxygen--rich; the optical depth is
extremely small during this phase, because of the small amount of silicate--dust in the 
envelope. After the carbon star stage is reached, $\tau_{10}$ increases because more and more carbon is accumulated at
the surface of the star, owing to the effects of TDU. In agreement with the discussion
in section \ref{dustagb}, the highest $\tau_{10}$ are reached by AGB stars of initial 
mass $\sim 2-2.5M_{\odot}$,
in the final evolutionary phases. The largest values of the optical depth, 
$\tau_{10} \sim 3$, are reached by the $Z=8\times 10^{-3}$ models; low--mass AGB stars
of $Z=10^{-3}$ evolve at optical depths below unity, whereas the $Z=4\times 10^{-3}$
models show an intermediate behaviour. This can be explained as follows: higher--Z
models evolve at lower effective temperatures, which favours dust formation, because the
region where gas molecules condense into dust is closer to the surface of the star, in a 
region of higher density.

The stars with initial mass above $3~M_{\odot}$ experience HBB and behave differently.
While in their counterparts of lower mass the optical depth depends strongly on 
the amount of carbon in the convective envelope, in massive AGB stars the dust formation 
process is mainly determined by the strength of HBB \citep{paperI}. The phase with the
highest infrared emission occurs during the phase of strongest HBB, when 
both luminosity and mass loss rate reach their maximum values. The luminosity in these
stars peaks in an intermediate evolutionary stage \citep[see Fig. 6 in][]{paperIV},
when $\tau_{10}$ is the highest shown. The tracks of the  $4~M_{\odot}$ and $5.5~M_{\odot}$ models in the three panels of Fig. \ref{ftau} reach the maximum $\tau_{10}$ during the largest luminosity phase. Inspection of Fig. \ref{ftau} confirms
that oxygen--rich models of higher mass have a stronger infrared emission, because the 
strength of HBB increases with the initial mass of the star; this holds independently
of the metallicity.

\begin{figure*}
\begin{minipage}{0.33\textwidth}
\resizebox{1.\hsize}{!}{\includegraphics{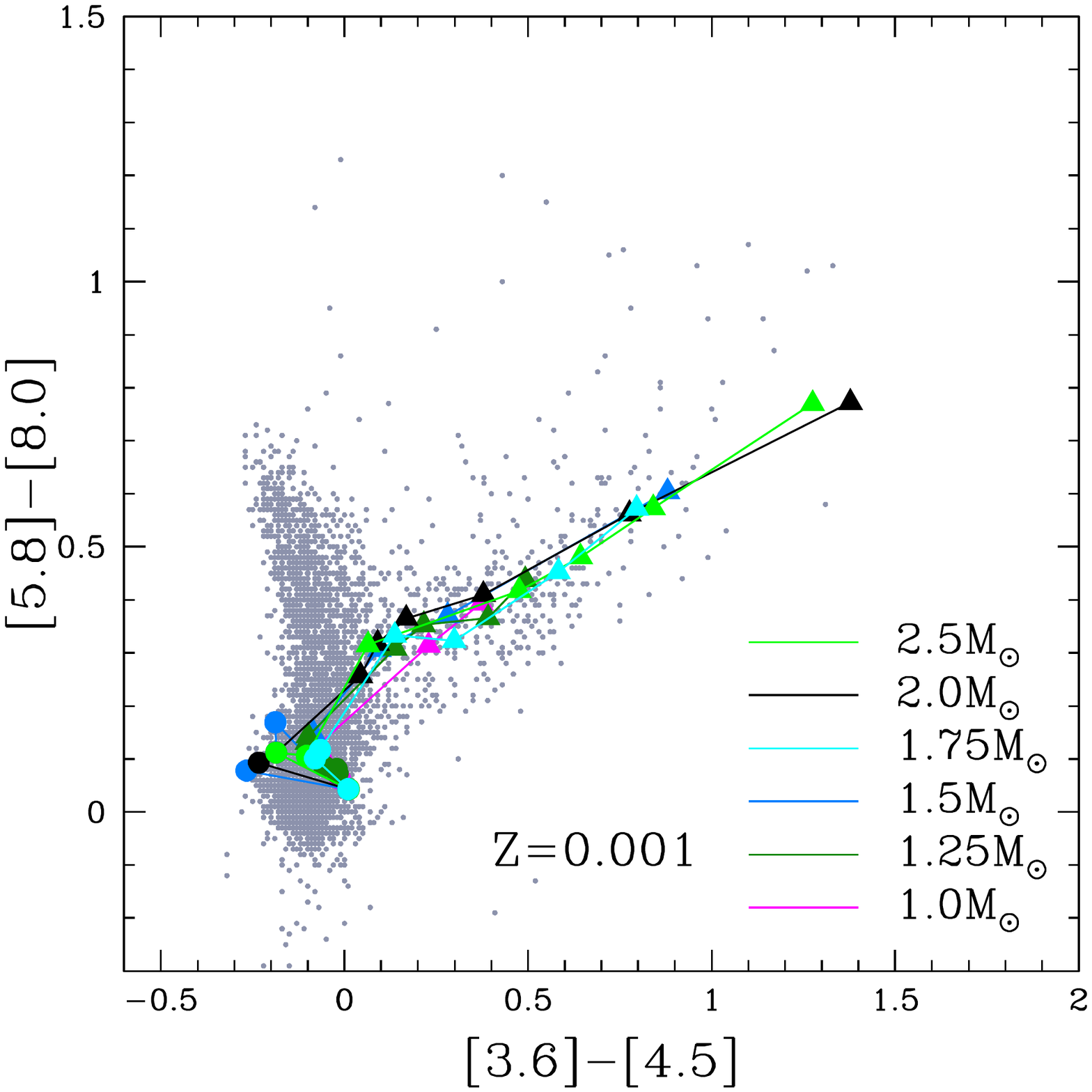}}
\end{minipage}
\begin{minipage}{0.33\textwidth}
\resizebox{1.\hsize}{!}{\includegraphics{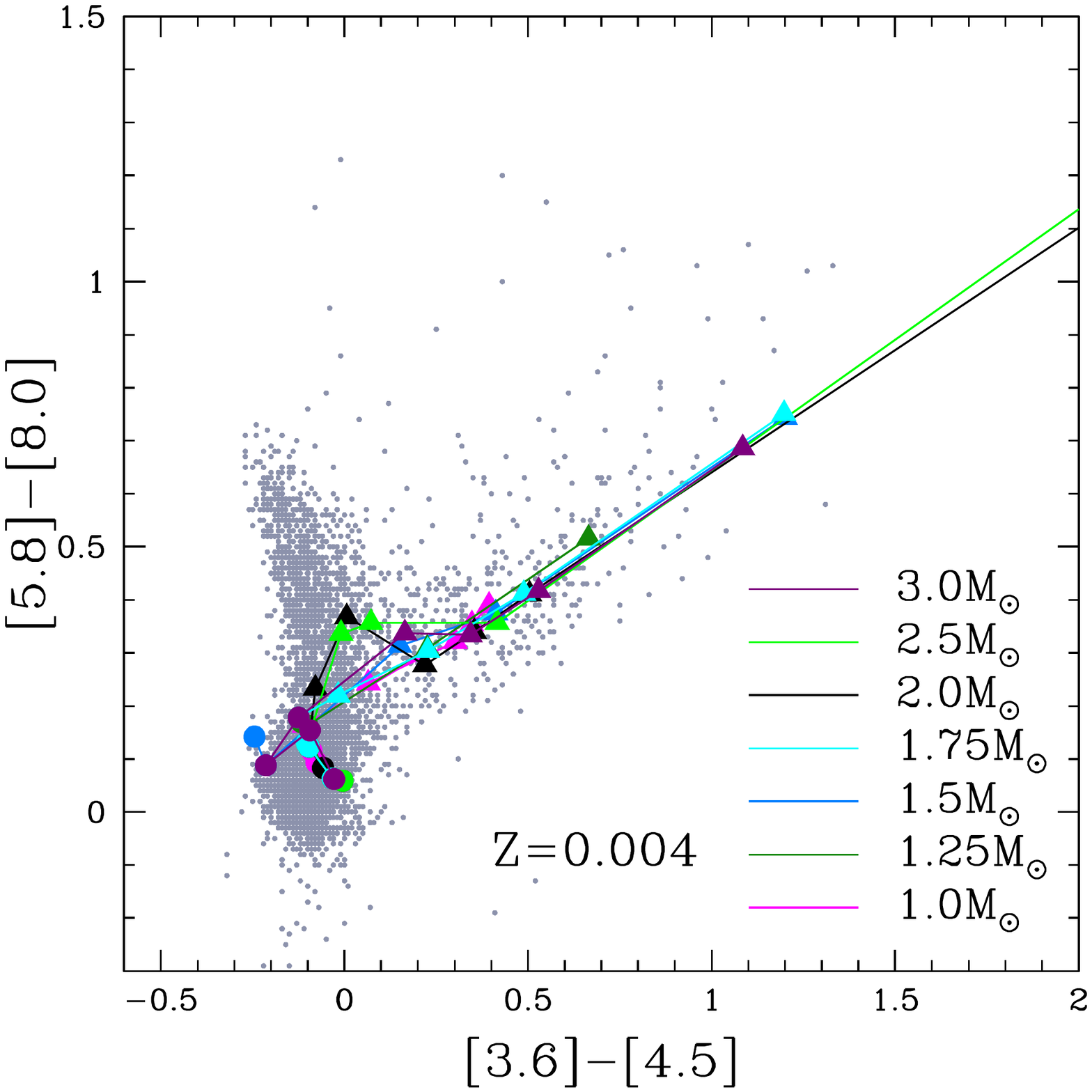}}
\end{minipage}
\begin{minipage}{0.33\textwidth}
\resizebox{1.\hsize}{!}{\includegraphics{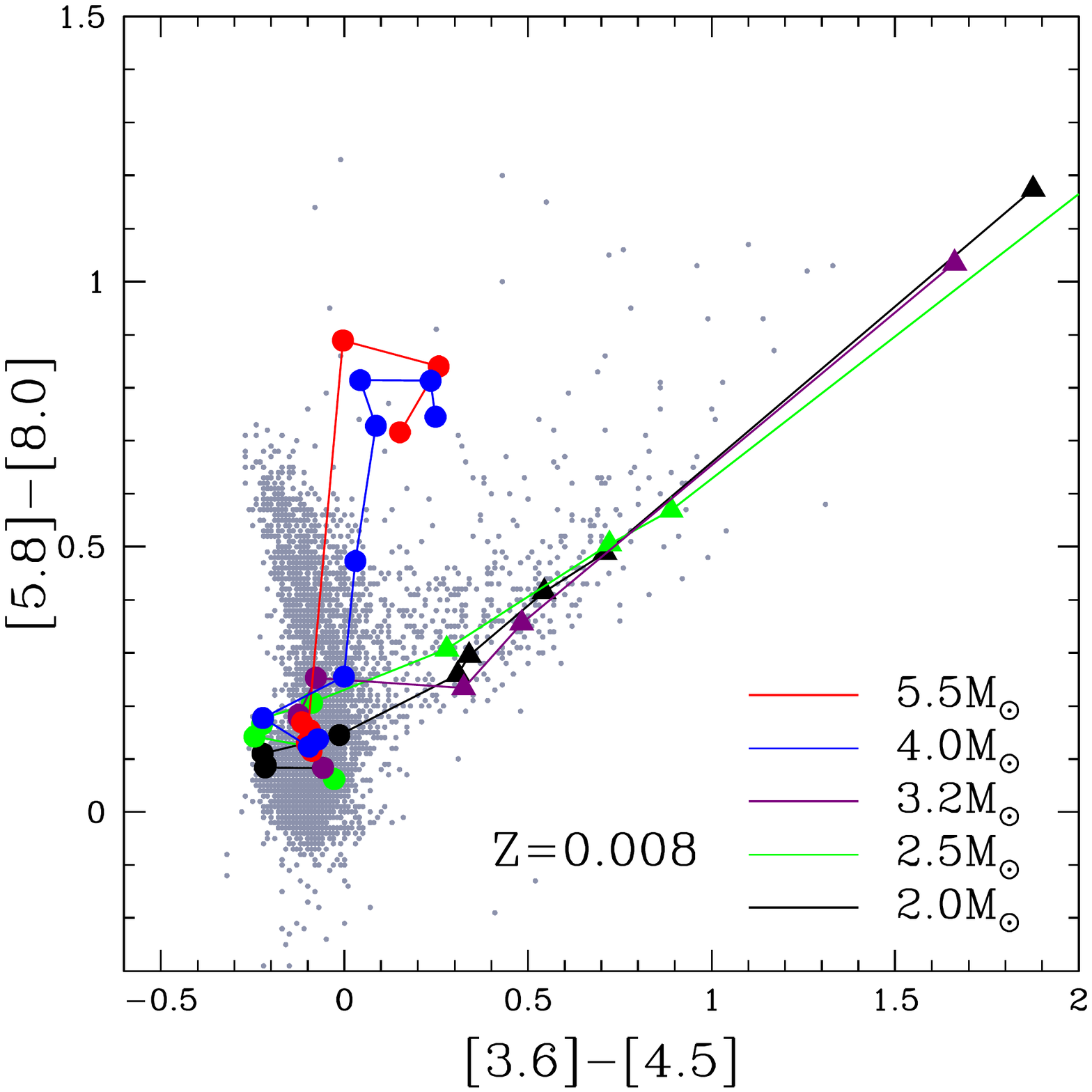}}
\end{minipage}
\vskip-20pt
\caption{Evolutionary tracks in the colour--colour ($[3.6]-[4.5], [5.8]-[8.0]$) plane of 
AGB stars of different initial mass and metallicity $Z=10^{-3}$ (left panel), 
$Z=4\times 10^{-3}$ (middle) and $Z=8\times 10^{-3}$ (right). The grey points represent
data of AGB stars in the SMC from \citet{boyer11}. Full circles indicate phases during which the
stars are oxygen--rich, whereas full triangles correspond to carbon--rich objects.
}
\label{ftracceccd}
\end{figure*}

The highest degree of obscuration reached in these models, $\tau_{10} \sim 1$, is smaller
than in their counterparts of smaller mass $\tau_{10} \sim 3$; this is due to the larger content of carbon in low--mass AGB stars compared to the silicon in the external regions of more massive AGB stars; an additional motivation is that the extinction coefficient of solid carbon dust is higher than the corresponding coefficients of silicates.

\begin{figure*}
\begin{minipage}{0.33\textwidth}
\resizebox{1.\hsize}{!}{\includegraphics{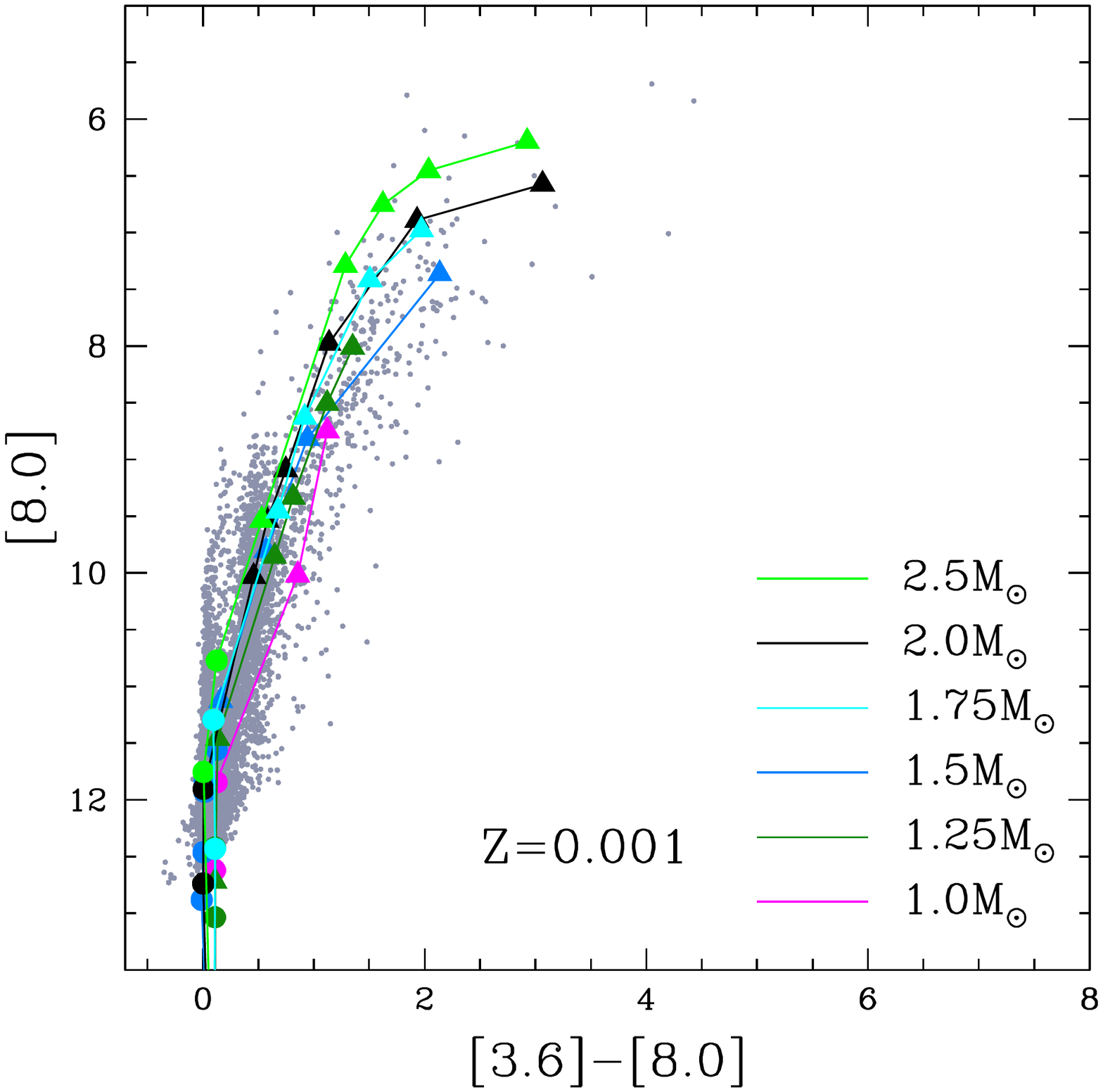}}
\end{minipage}
\begin{minipage}{0.33\textwidth}
\resizebox{1.\hsize}{!}{\includegraphics{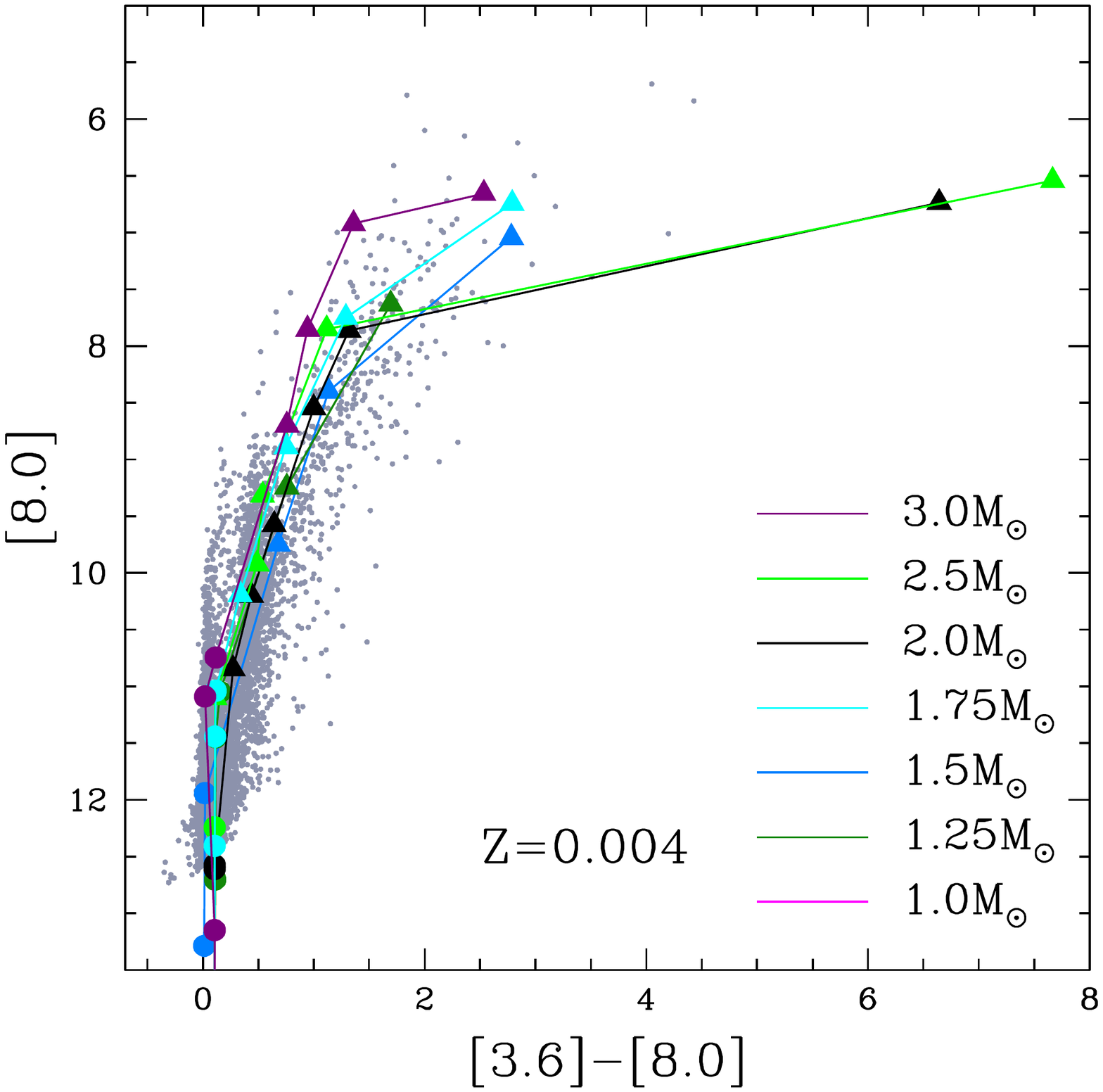}}
\end{minipage}
\begin{minipage}{0.33\textwidth}
\resizebox{1.\hsize}{!}{\includegraphics{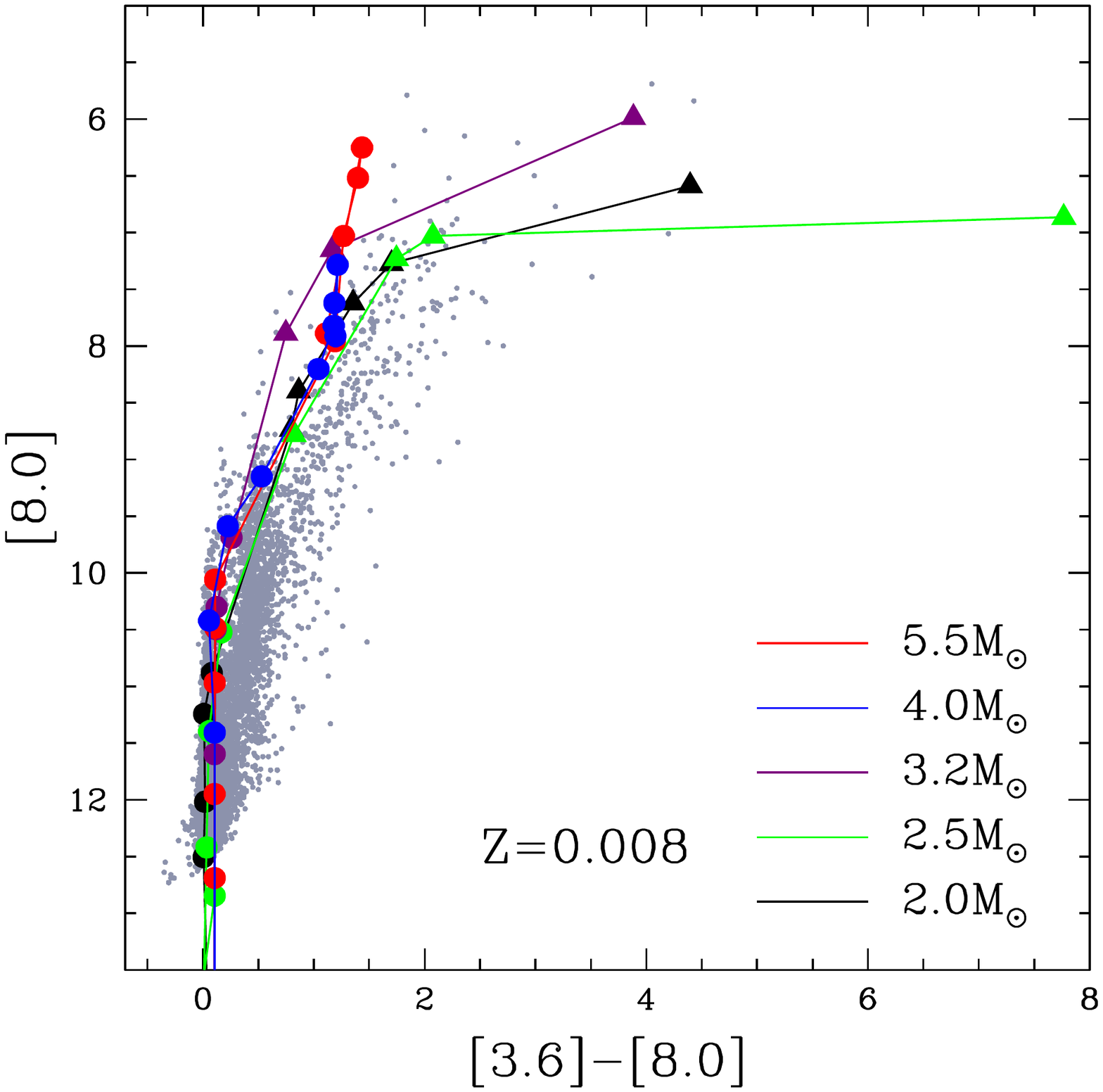}}
\end{minipage}
\vskip-20pt
\caption{Evolutionary tracks in the colour--magnitude ($[3.6]-[8.0], [8.0]$) plane of 
AGB stars of different initial mass and metallicity. The meaning of the symbols is the
same as in Fig.~\ref{ftracceccd}.
}
\label{ftraccecmd}
\end{figure*}

\subsection{The infrared colours of AGB stars}
\begin{figure}
\resizebox{1.\hsize}{!}{\includegraphics{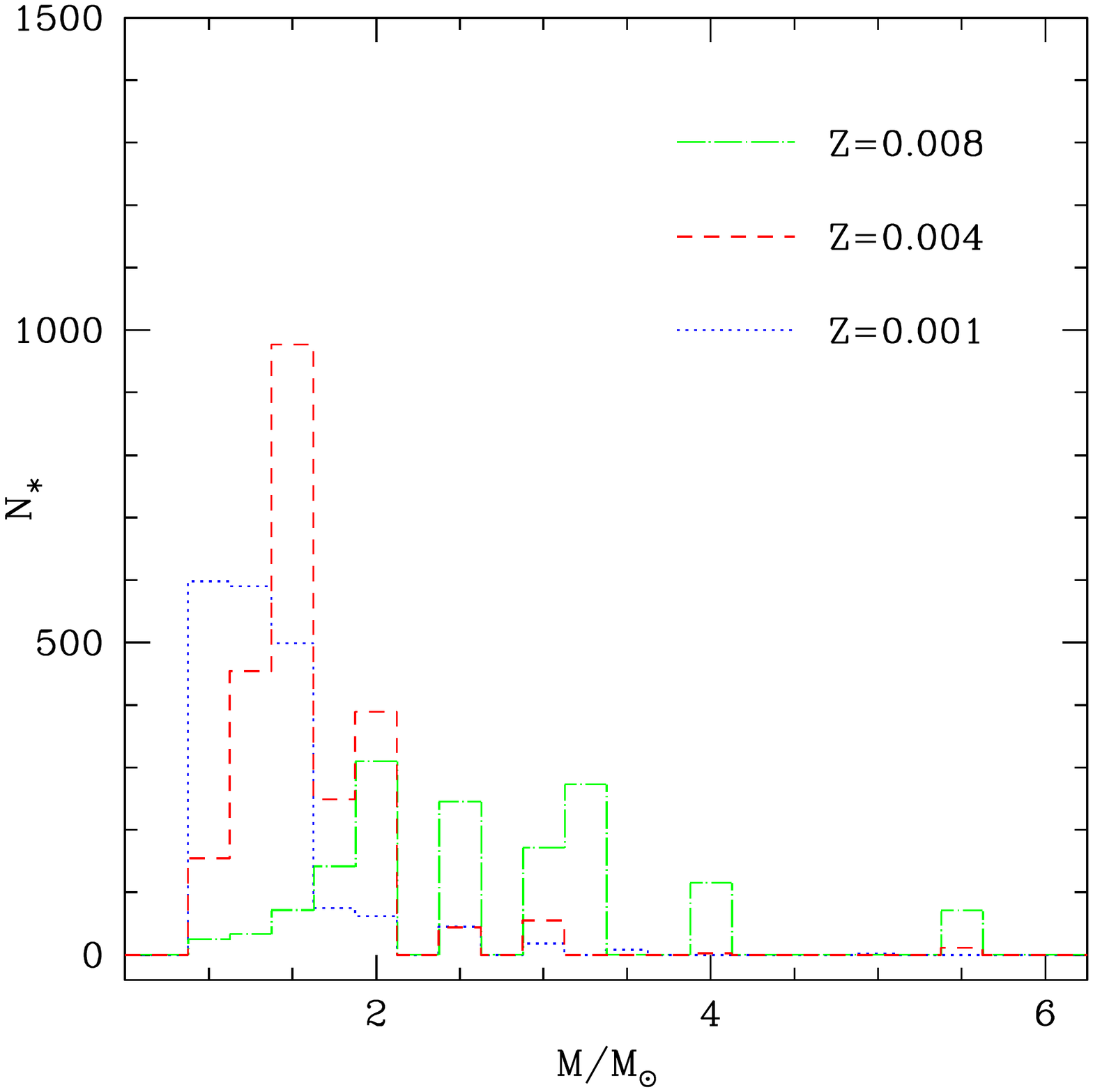}}
\vskip-50pt
\caption{The mass distribution of the AGB population of the SMC, according to our
synthetic modelling. The various masses are separated among  
the three different metallicities used in the present analysis: $Z=10^{-3}$ (dotted, blue
histogram), $Z=4\times 10^{-3}$ (dashed, red), $Z=8\times 10^{-3}$ (dotted--dashed, green).
}
\label{fisto}
\end{figure}

Fig. \ref{ftracceccd} and \ref{ftraccecmd} show the evolutionary tracks of AGB models
of various initial mass in the colour--colour ($[3.6]-[4.5]$, $[5.8]-[8.0]$) diagram
(hereinafter CCD) and in the colour--magnitude ($[3.6]-[8.0]$, $[8.0]$) plane (CMD).
For clarity reasons, we show some stellar models which, according to the star 
formation history and the age--metallicity relationship of the SMC (see Fig. \ref{fsfh}), will dominate the predicted synthetic population (see Fig. \ref{fisto}).
In the same figures we show the observations from \citet{boyer11}.

As shown in Fig. \ref{ftau}, stars with mass $1~M_{\odot} < M_{init} < 3~M_{\odot}$ evolve at larger and
larger optical depths as their envelope becomes more enriched in carbon. The tracks evolve
to redder infrared colours, owing to the presence of carbonaceous dust particles in
the surroundings of the star. In the CCD the models define an evolutionary line of constant
slope. Only models of mass $M \geq 2~M_{\odot}$ evolve at $[3.6]-[4.5] > 1.5$, because
they accumulate more carbon in the external regions, due to the higher number of thermal
pulses experienced. The models of low metallicity do not reach such red colours; this is
an effect of their higher effective temperature, as discussed in section \ref{irspectra}.
In the CMD the main effect of the gradual carbon enrichment in the external regions of the
stars is the rightwards excursion of the tracks, towards redder $[3.6]-[8.0]$ colours.
In this plane the threshold colour separating the $Z=10^{-3}$ models from their more
metal rich counterparts is $[3.6]-[8.0] \sim 3$.

The stars with mass $M_{init}  > 3~M_{\odot}$ experience HBB and produce mainly silicates, with
traces of alumina dust. During the maximum luminosity phase the stars reach their 
reddest IR colours. The infrared emission during these phases is sensitive to mass and metallicity, according to the discussion in the previous section (see Fig. \ref{ftau}). In the CCD
the evolutionary tracks (see right panel of Fig. \ref{ftracceccd}) evolve to the region
at $[3.6]-[4.5] \sim 0.2$, $[5.8]-[8.0] \sim 0.8$. The tracks of these obscured AGB stars
bifurcate from their counterparts of lower mass, owing to the presence of the silicates
feature at $9.7 \mu$m, affecting the $8.0\mu$m flux. A bifurcation among the C--rich and
the oxygen--rich tracks is also found in the CMD (see right panel of Fig. \ref{ftraccecmd}),
the latter evolving along a more vertical sequence.

\section{AGB stars in the SMC: understanding the IR colours}
\label{intepret}
To interpret the {\it Spitzer} observations of AGB stars in the SMC, we produced synthetic
diagrams in the CCD and CMD planes, with the same methods described in D15.
For each epoch, we extracted a number of stars, proportional to the SFR by \citet{harris04},
shown in Fig. \ref{fsfh}. The mass distribution follows a standard Salpeter IMF with index
$x=-1$, whereas the number of extractions for each mass scales with the overall duration of
the AGB phase.

The mass and metallicity distribution of the stars extracted is shown in
Fig. \ref{fisto}. In the mass domain of the stars experiencing HBB, $M_{init} > 3~M_{\odot}$,
we note two peaks, at $M\sim 4, 5.5~M_{\odot}$; the first
corresponds to the peak in the SFH which occurred 150Myr ago (see Fig. \ref{fsfh}), whereas
the maximum at $M\sim 5.5~M_{\odot}$ concerns younger epochs, $\sim 90$Myr ago.
These stars belong to the more metal--rich population, with metallicity $Z=8\times 10^{-3}$,
which provided the greatest contribution during epochs younger than $\sim 1$Gyr.
In the low--mass domain the dominant AGB population is
given by $Z=4\times 10^{-3}$ stars, with also a significant contribution from the
low--metallicity component. The peak in the mass distribution around $M=1.5~M_{\odot}$
correspond to the peak in the SFH which occurred $\sim 2.5$Gyr ago (see Fig. \ref{fsfh}).

The comparison among the observations of SMC AGB stars and the results from our synthetic 
modelling are shown in Figs. \ref{fsmc_ccd} and \ref{fsmc_cmd}. In the two figures we compare 
the observed (left panels) and expected (right panels) distribution of AGB stars in the CCD (Fig. \ref{fsmc_ccd}) 
and CMD (Fig. \ref{fsmc_cmd}) planes. The samples of spectroscopically confirmed AGB stars, described in section \ref{obs}, are also shown in the same figures. Fig. \ref{fsmc_ccd_met} and \ref{fsmc_cmd_met} also show the comparison among the observation and theoretical predictions in the CCD and in CMD planes; in this case, however, the two panels show the distribution of carbon--rich stars (right) and oxygen--rich stars (left), divided among the various metallicity components.

\subsection{Unobscured stars}
\begin{figure*}
\begin{minipage}{0.49\textwidth}
\resizebox{1.\hsize}{!}{\includegraphics{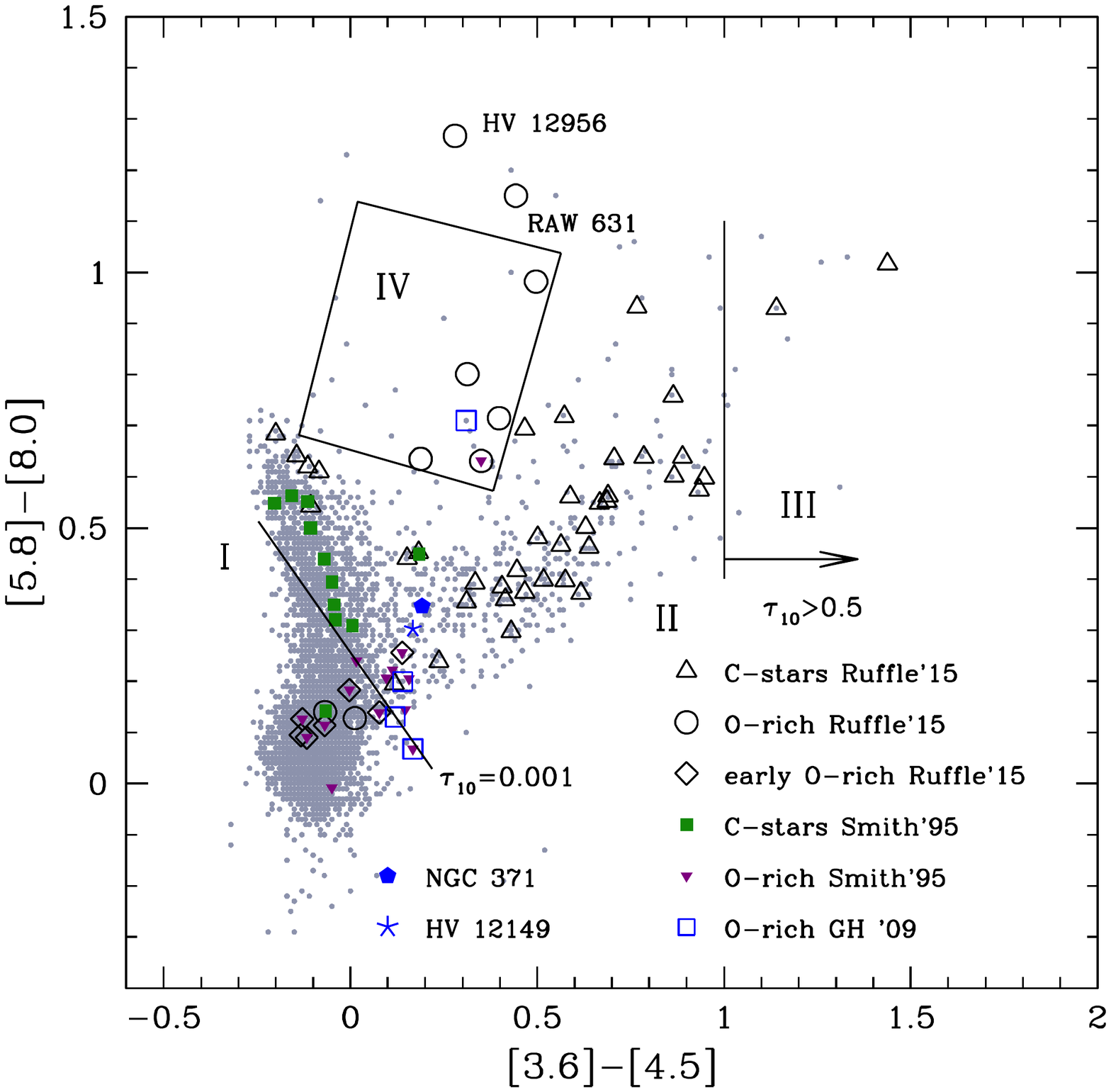}}
\end{minipage}
\begin{minipage}{0.49\textwidth}
\resizebox{1.\hsize}{!}{\includegraphics{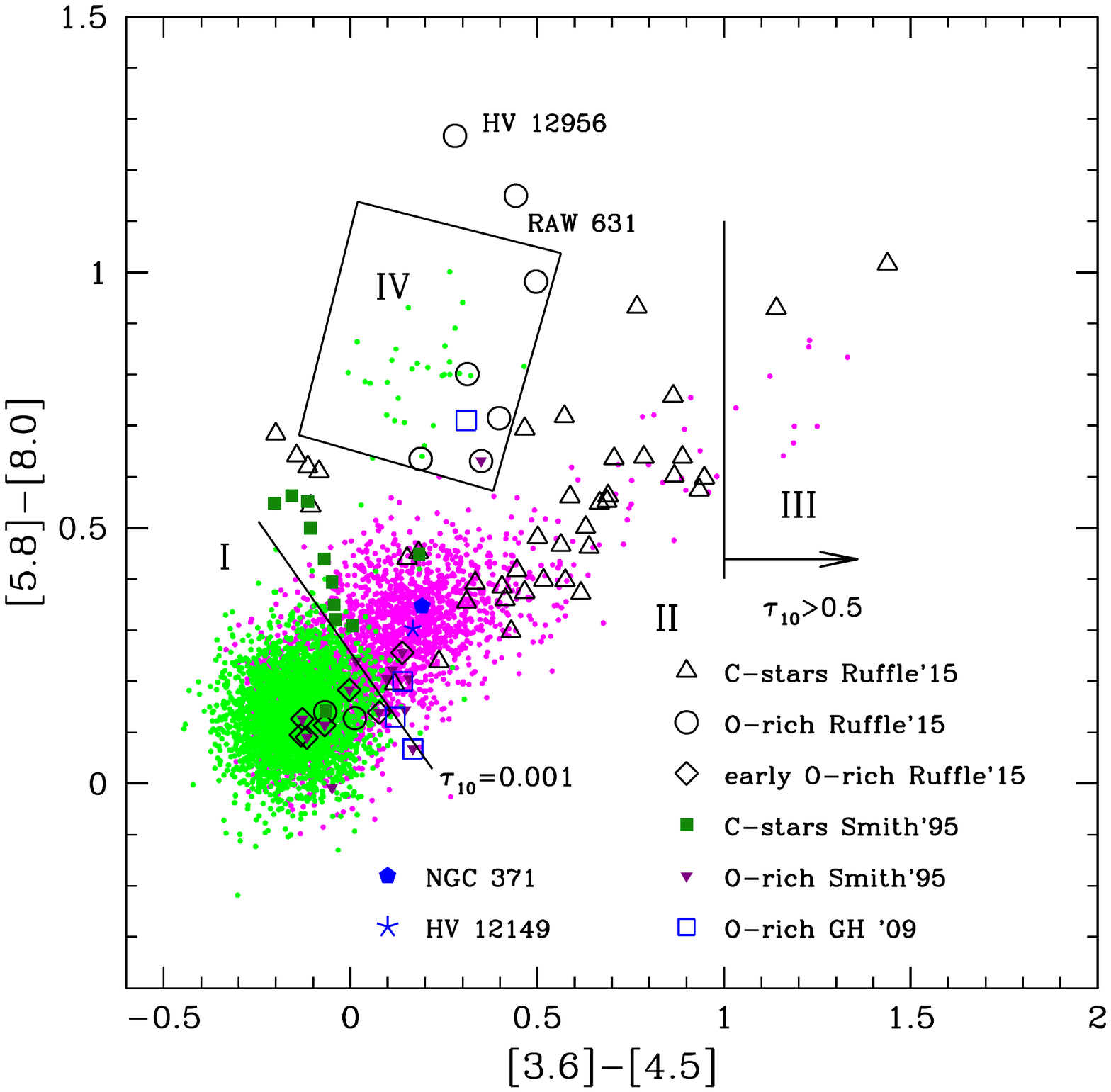}}
\end{minipage}
\vskip-50pt
\caption{Left: the distribution of the AGB sample of the SMC by \citet{boyer11} in
the colour--colour ($[3.6]-[4.5], [5.8]-[8.0]$) plane. We show several spectroscopically 
confirmed samples: \citet{ruffle15}, separated into carbon stars (black open triangles), oxygen--rich objects (black open circle) and early oxygen--rich objects (black open diamond), carbon stars
(green full square) and oxygen--rich objects (violet full reversed triangle) by \citet{smith95} and oxygen--rich stars (blue open square) from \citet{anibal09}. NGC 371 (Blue full diamond) and HV 12149 (blue tar) are discussed in sec. \ref{hbbagb}. Right: the results from our synthetic modelling in the same plane shown in
the left panel. Stars expected to be oxygen--rich are shown in green, whereas
carbon stars are indicated in magenta. The spectroscopically 
confirmed samples by \citet{ruffle15}, \citet{smith95} and \citet{anibal09} are also shown. The open square and the full triangle within region IV indicate, respectively, stars HV1375 and IRAS00483-7347, discussed in section 5.3.
The lines in the figure
delimit regions I, II, III, IV, populated, respectively, by scarcely obscured AGB stars,
carbon stars, extremely obscured carbon stars and stars experiencing Hot Bottom Burning.
}
\label{fsmc_ccd}
\end{figure*}

\begin{figure*}
\begin{minipage}{0.49\textwidth}
\resizebox{1.\hsize}{!}{\includegraphics{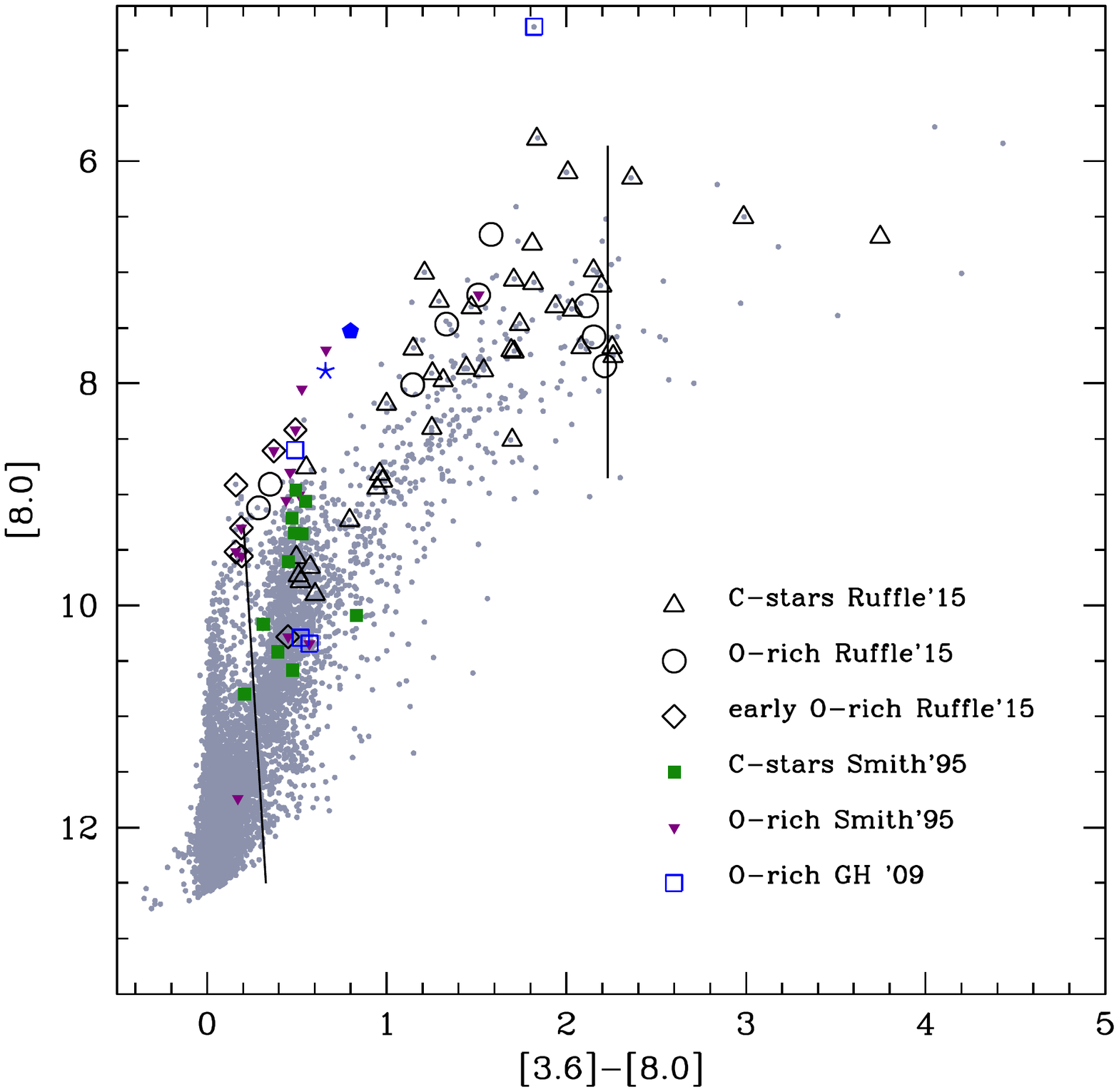}}
\end{minipage}
\begin{minipage}{0.49\textwidth}
\resizebox{1.\hsize}{!}{\includegraphics{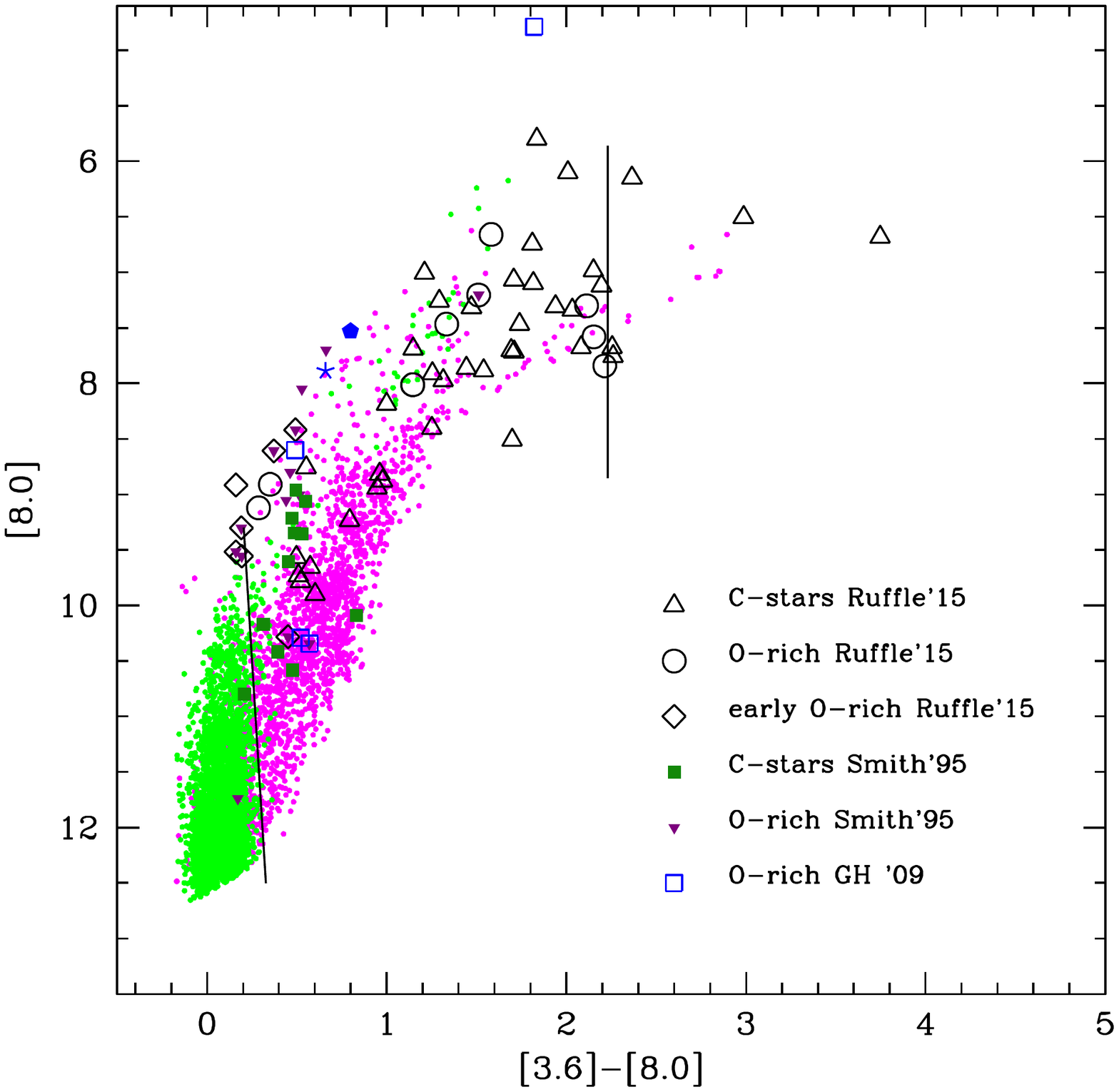}}
\end{minipage}
\vskip-50pt
\caption{The same as Fig.~\ref{fsmc_ccd}, but referred to the colour--magnitude
($[3.6]-[8.0], [8.0]$) plane. The line on the lower--left side of the diagram
separates scarcely obscured objects (left) from C--stars (right), whereas the
line at $[3.6]-[8.0] \sim 2.2$ corresponds to the line on the right side of
the CCD shown in Fig.~\ref{fsmc_ccd}. The spectroscopically confirmed samples by 
\citet{ruffle15}, \citet{anibal09} and \citet{smith95} are shown, with the same symbols as in Fig.~\ref{fsmc_ccd}.
}
\label{fsmc_cmd}
\end{figure*}

The region of the CCD
centered at $[3.6]-[4.5] \sim -0.1$, $[5.8]-[8.0] \sim +0.1$ is populated by AGB stars with an 
extremely small degree of obscuration, whose spectrum is not expected to show significant 
dust features. The red border of this region, (hereinafter region I), is represented
by the diagonal line in Fig. \ref{fsmc_ccd}, indicating stars with $\tau_{10} \sim 0.001$;
this choice is such that almost the totality of the oxygen--rich objects evolve bluewards of this line in the CCD (see Fig. \ref{fsmc_ccd_met}). As shown in Fig. \ref{ftracceccd}, the only exception here is represented by stars of initial mass above $\sim 3~M_{\odot}$, whose tracks cross the afore mentioned line when HBB begins.

According to our interpretation, the majority ($\sim 67\%$) of AGB stars in region I are
low--mass, oxygen--rich stars, in the first part of the AGB phase, before the C--star
stage is reached. They are distributed among $Z = 4\times 10^{-3}$ stars of mass 
$1-2~M_{\odot}$ ($\sim 27\%$), $Z = 8\times 10^{-3}$ AGB stars of mass $2-3~M_{\odot}$ ($\sim 23\%$),
$Z = 10^{-3}$ objects, of mass below $2~M_{\odot}$ ($\sim 17\%$).

A low fraction ($\sim 4\%$) of stars in region I of the CCD is composed of 
$Z = 8\times 10^{-3}$ AGB stars of higher mass, in the phases previous to
ignition of HBB; as discussed previously, and shown in Fig.~\ref{fisto}, the mass 
distribution of these stars peaks at $M \sim 4~M_{\odot}$ and $M \sim 5.5~M_{\odot}$. 
Scarcely obscured objects, classified as oxygen--rich stars in the sample by \citet{ruffle15} and \citet{smith95}, occupy this region of the CCD.
The oxygen--rich stars presented by \citet{smith95} are divided among AGB stars with or without lithium. Unfortunately, the detection of lithium is of little help in the present analysis, owing to the peculiar behavior of this fragile element during the AGB evolution. Lithium is heavily destroyed  by proton capture as soon as the AGB phase begins and is produced in great quantities when the temperature at the bottom of the surface convective region reaches $\sim 40MK$ \citep{cameron71}. The lithium--rich phase lasts until some $^3He$ is available in the envelope \citep{sackmam92}, thus it is limited to the initial part of the HBB phase \citep{mazzitelli99}.
Based on these arguments, it is not surprising that some of the stars in the \citet{smith95} sample present evidence of lithium in their spectra; for similar reasons we understand that some of the oxygen--rich stars studied by \citet{smith95} with a large infrared emission (see section \ref{hbbagb} below) show no signature of lithium in their spectra.

As shown in the right panel of Fig. \ref{fsmc_ccd_met}, according to our modelling, $\sim 29\%$ of the sources populating region I
are carbon stars, either at the very beginning 
of the C--star evolution, or in the evolutionary phases immediately following the ignition
of each thermal pulse; these objects have a mixed metallicity: half of them are 
$Z = 4\times 10^{-3}$ AGB stars of mass $M<2~M_{\odot}$, the remaining have metallicity
$Z = 10^{-3}$ and mass below $M<1.5~M_{\odot}$.
A word of caution is needed here. The comparison among the observations
and our synthetic model shows the presence of a group of stars, in the region at 
$[3.6]-[4.5] \sim -0.2$, $[5.8]-[8.0] \sim +0.7$, not reproduced by the models. We identify 
these stars as carbon--rich objects, not heavily obscured. 
 The synthetic spectrum of these AGB stars is determined by the spectral energy distribution from the 
central object, substantially unchanged by the optically thin envelope. The blue $[3.6]-[4.5]$ 
colours of these stars are not reproduced by the GRAMS atmosphere models used to produce
the synthetic SED. This same problem was already addressed by D15, and is
discussed in details by Srinivasan et al. (2011) (section 4.2.5). CO and $C_{3}$ absorption bands could be a possible explanation of this discrepancy.
We therefore suggest that part of the unobscured C--stars, which in our modelling
evolve into region I of the CCD, would indeed populate the region
in the higher portion of the CCD, currently uncovered by the models.
C-stars in the \citet{smith95} sample populate this region of the color--colour plane.

In the CMD, the group of scarcely obscured AGB stars populate the lower--left region.
The $[8.0]>12$ mag zone (see left panel of Fig.~\ref{fsmc_cmd_met}) is composed of
stars of various surface chemistry. The sequences of C--stars and oxygen--rich AGB stars 
begin to bifurcate for $[3.6]-[8.0]\sim 0.3$ and $[8.0]<12$ mag (see left panel of Fig.~\ref{ftraccecmd}): oxygen--rich stars trace a more vertical sequence, 
with $[3.6]-[8.0]$ colours bluer than $\sim 0.3$, whereas carbon stars populate 
the region $[3.6]-[8.0] > 0.3$. This effects, discussed in D15, originates 
from the silicate feature at $\sim 9.7\mu$m, present in the spectra of dusty, oxygen--rich 
AGB stars: for a given $[3.6]-[8.0]$ colour, the latter stars are brighter in the $8.0\mu$m band,
compared to their carbon--rich counterparts.

For what concerns oxygen--rich objects, the analysis of the distribution of the stars in 
the CMD, compared to the CCD, offers a better opportunity to disentangle stars of
different metallicity; this is clearly shown in the left panel of Fig. \ref{fboyer}. While for magnitudes $[8.0]>11.5$ most of the stars belong to
the population of lower metallicity, in the 11 mag $< [8.0] < 11.5$ mag region we mainly find
$Z = 4\times 10^{-3}$ objects, while for magnitudes $[8.0]<11$ we practically have only
stars of metallicity $Z = 8\times 10^{-3}$. This is because the surface regions of 
higher metallicity models contain more silicon and aluminum, which favour the formation
of larger amounts of silicates and alumina dust, and a higher degree of obscuration of
the radiation from the central stars, hence a higher flux in the $8.0\mu$m band. An 
additional motivation for this behaviour is that low--mass, higher metallicity AGB stars
evolve longer as oxygen--rich objects, compared to their counterparts of lower Z
(see Fig.~\ref{ftau}); while the evolutionary tracks of the $Z = 8\times 10^{-3}$
(and, to a lower extent, of the $Z = 4\times 10^{-3}$) models follow the almost vertical
sequence traced by oxygen--rich stars for the majority of the AGB life, the corresponding
tracks of the $Z = 10^{-3}$ models turn earlier (at higher $[8.0]$ magnitudes)
towards redder $[3.6]-[8.0]$ colours.

\subsection{Carbon stars}
\label{cstars}
In the CCD, carbon stars populate a diagonal strip of constant slope, from the edge of region
I towards red infrared colours. This is shown in Fig.~\ref{ftracceccd}, where the 
evolutionary tracks of low--mass AGB stars overlap with the observations by \citet{boyer11}.

Similarly to D15, we interpret the AGB stars in region II of the CCD as an obscuration sequence
of carbon stars: the objects with the largest infrared emission are those with the 
higher surface carbon content, which experienced a higher number of TDU episodes. The
optical depth increases towards the red side of the sequence, owing to the large quantities
of dust (mainly solid carbon grains) in the surroundings of these stars. 

\begin{figure*}
\begin{minipage}{0.49\textwidth}
\resizebox{1.\hsize}{!}{\includegraphics{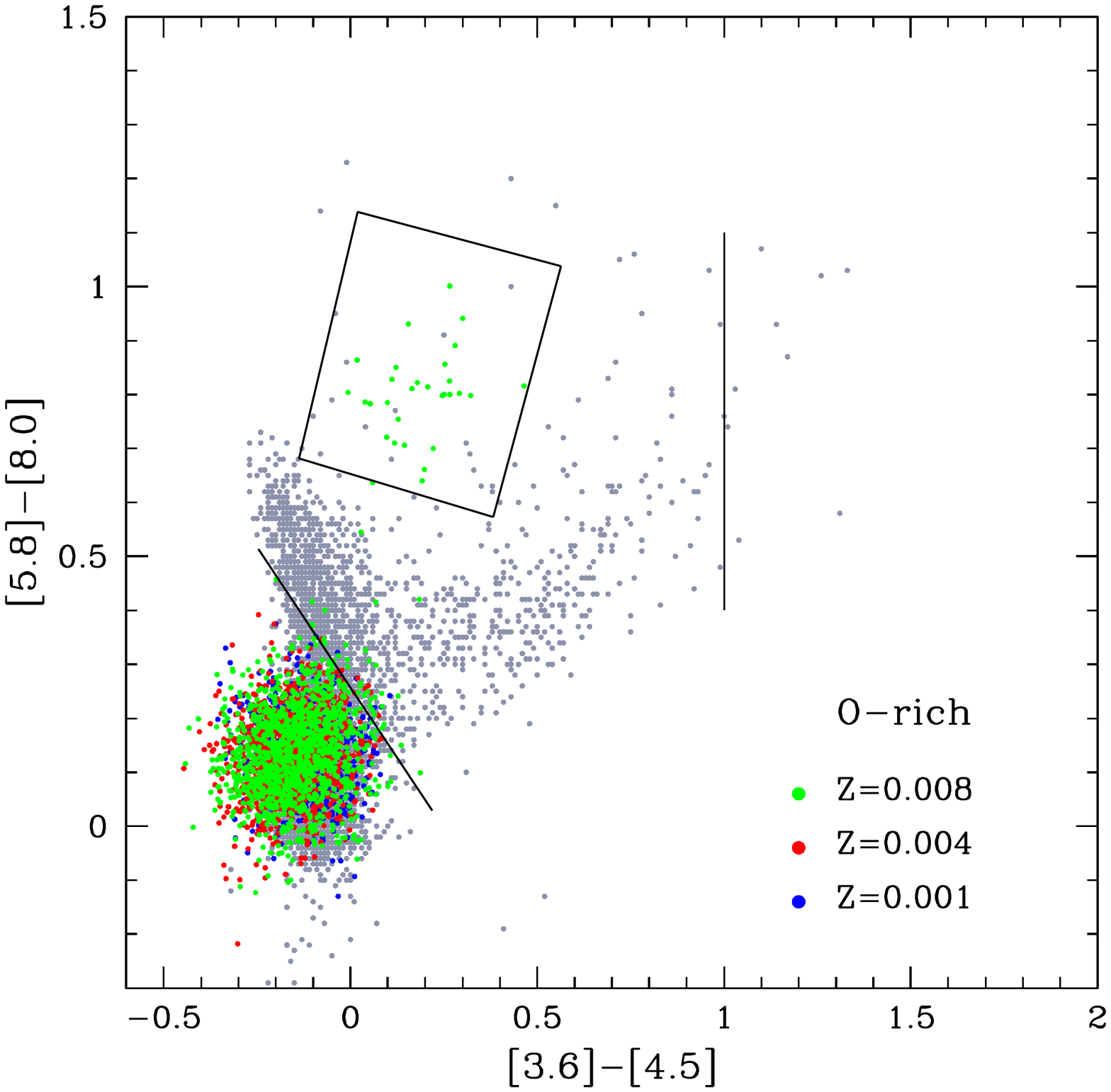}}
\end{minipage}
\begin{minipage}{0.49\textwidth}
\resizebox{1.\hsize}{!}{\includegraphics{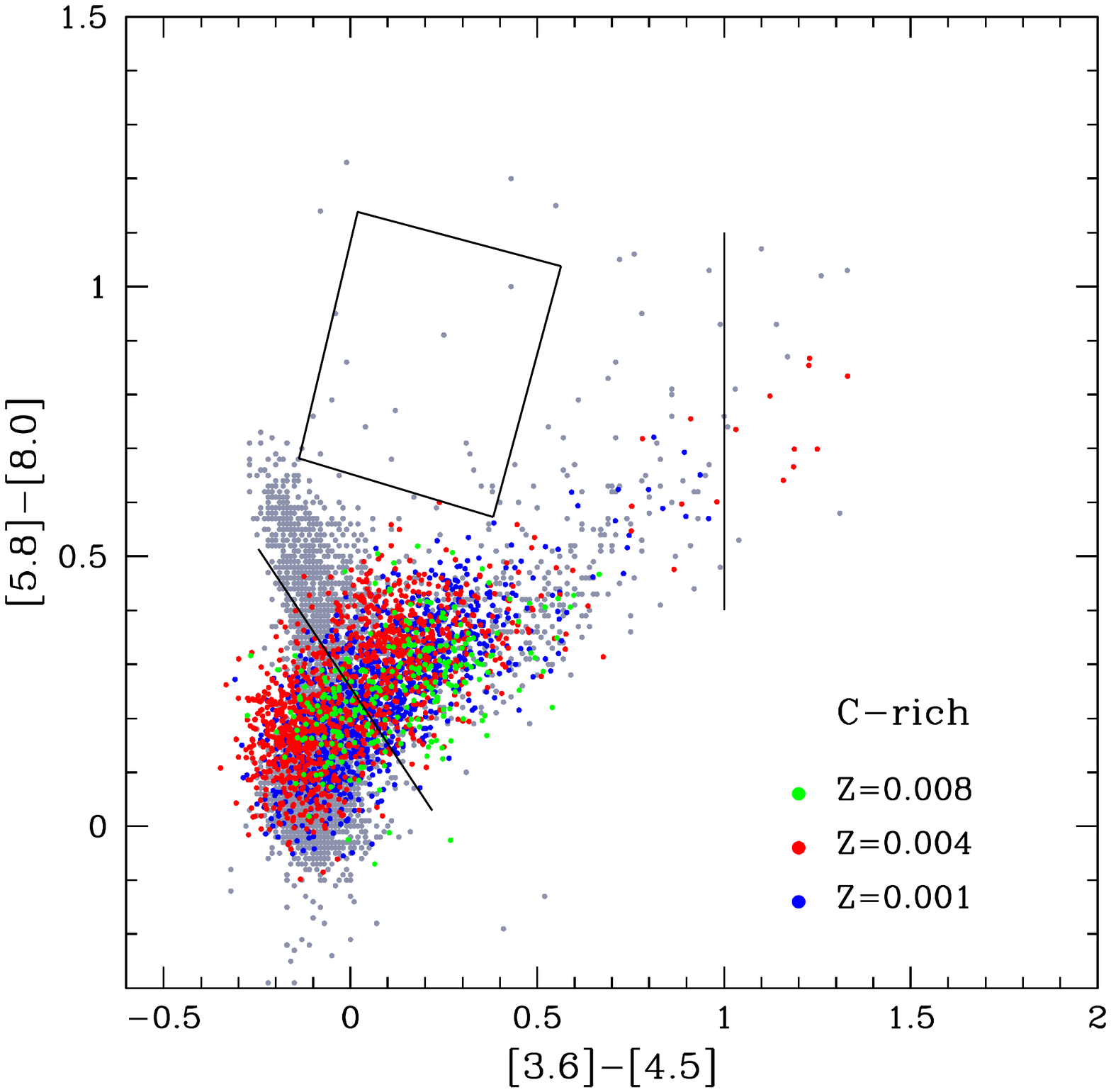}}
\end{minipage}
\vskip-50pt
\caption{Left: the expected distribution of oxygen--rich stars from our synthetic modelling in
the colour--colour ($[3.6]-[4.5], [5.8]-[8.0]$) plane. We show the three metallicity components in different colours: $Z=8\times 10^{-3}$ in green, $Z=4\times 10^{-3}$ in red and $Z=10^{-3}$ in blue. The observed AGB sample from \citet{boyer11} is also shown in grey. Right: the expected distribution of carbon stars from our synthetic modelling in the colour--colour, with the same colour coding as in the left panel.}
\label{fsmc_ccd_met}
\end{figure*}

\begin{figure*}
\begin{minipage}{0.49\textwidth}
\resizebox{1.\hsize}{!}{\includegraphics{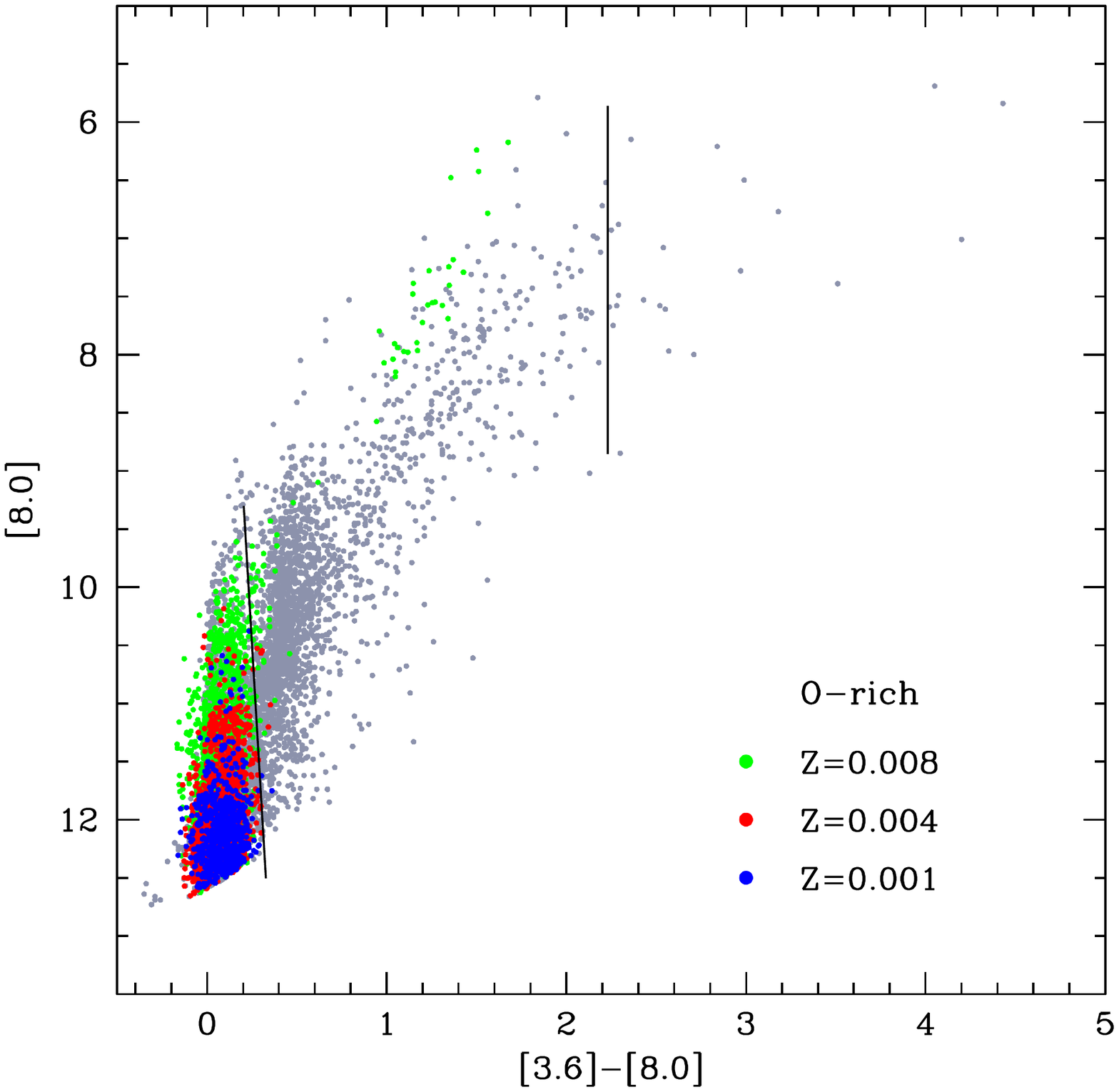}}
\end{minipage}
\begin{minipage}{0.49\textwidth}
\resizebox{1.\hsize}{!}{\includegraphics{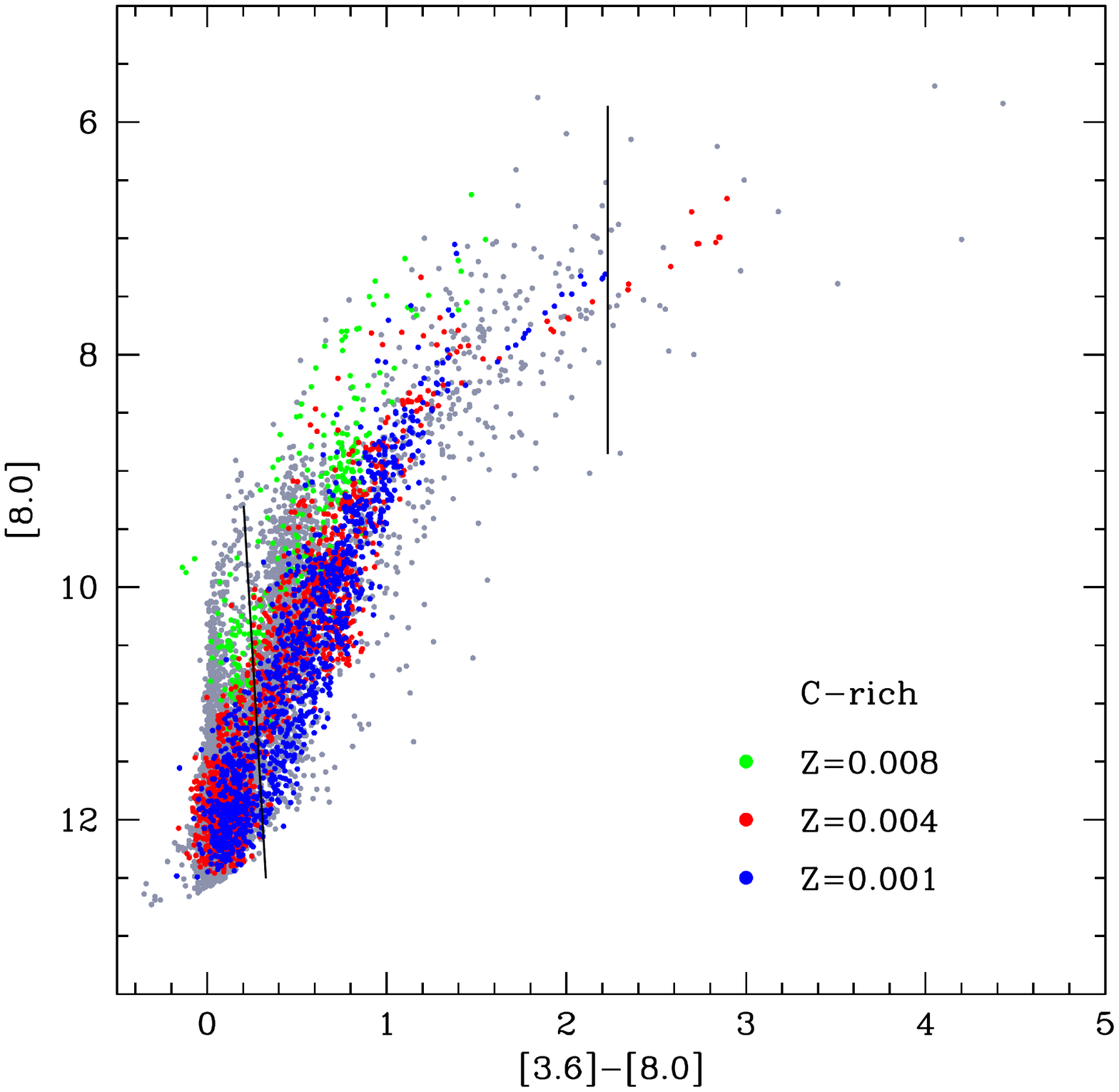}}
\end{minipage}
\vskip-50pt
\caption{The same as Fig. \ref{fsmc_ccd_met}, but referred to 
the colour--magnitude ($[3.6]-[8.0],[8.0]$) plane.}
\label{fsmc_cmd_met}
\end{figure*}

As stated previously, the blue border of region II was chosen in such a way that this
regions is almost exclusively populated by carbon stars; however, a small fraction of
C--rich objects is also expected to populate region I in the CCD (see right panel of Fig. \ref{fsmc_ccd_met}).

The results shown in Fig.~\ref{fsmc_ccd} indicate a satisfactory agreement among the 
observed and expected distribution of AGB stars in this region of the CCD; the colours of the 
theoretical models nicely fit the position of the carbon AGB stars sample by \citet{ruffle15}. 

Our analysis indicate that the carbon star sample in the SMC
is composed of objects of metallicity $Z = 4\times 10^{-3}$ of mass 
$1.5~M_{\odot} \leq M \leq 2~M_{\odot}$, formed between $700$Myr and 
$1.5$Gyr ago and, in equal part, by low--mass ($M < 1.5~M_{\odot}$) stars belonging to the $Z = 10^{-3}$ 
stellar component, $1.5-5$Gyr old. Only 8\% of the carbon stars are objects of metallicity $Z = 8\times 10^{-3}$ of mass 
$1.5~M_{\odot} < M < 3.5~M_{\odot}$, formed between $300$Myr and 1.7Gyr.

The stars with the largest infrared emission, with $[3.6]-[4.5] > 1$, populating
region III in the CCD, descend from stars of metallicity $Z = 4\times 10^{-3}$ (see right panel of 
Fig. \ref{fsmc_ccd_met}) and initial mass $M \sim 1.5~M_{\odot}$; according to our interpretation, these highly 
obscured objects formed $\sim 1.5$Gyr ago. The dust in their circumstellar envelope is 
mainly composed of solid carbon particles with size $\sim 0.2\mu$m; the optical depth of 
these stars is $\tau_{10}>0.5$. Fig.~\ref{fsmc_ccd} shows that the number of stars in this 
region of the CCD decreases as the colours become redder. This is partly because only the 
$Z = 4\times 10^{-3}$ stars reach the zone populated by the most obscured AGB stars; an
additional reason is that the stars with the largest infrared emission lose more rapidly
their external envelope, which makes the remaining evolutionary phases shorter
(see Fig.~3 in D15). The two carbon stars in the \citet{ruffle15} sample with the largest 
infrared emission belong to this group.

In the CMD plane, obscured carbon stars populate the region extending from
$[3.6]-[8.0] \sim 0.3$ to $[3.6]-[8.0] \sim 4$. The group of stars in the region
included between the two lines in the CMD (see Fig.~\ref{fsmc_cmd}) is interpreted
as composed mainly of carbon stars, with a small contribution from obscured, oxygen--rich
AGB stars (see next section). The vertical line at $[3.6]-[8.0] \sim 2.2$ delimits the region
where the most obscured C--stars, populating region III in the CCD, evolve. 

The position of the C--rich stars in the \citet{ruffle15} sample is nicely reproduced
also in this plane.

\subsection{AGB stars experiencing Hot Bottom Burning}
\label{hbbagb}
In the observed CCD of the SMC, (see Fig.~\ref{fsmc_ccd}), we see 
a few stars within region IV, with colours 
$[3.6]-[4.5] \sim 0.1-0.3$, $[5.8]-[8.0] \sim 0.5-1$. According to our interpretation,
based on the evolutionary tracks shown in the right panel of Fig.~\ref{ftracceccd}, these 
are the descendants of stars formed $100-400$Myr ago, with initial mass 
peaking at $4 M_{\odot}$ and $5.5 M_{\odot}$, currently experiencing HBB. This group correspond 
to the AGB stars defined as ``HBBS" in D15 and it is composed only by stars with metallicity  $Z=8\times 10^{-3}$, as shown in the left panel of 
Fig. \ref{fsmc_ccd_met}. 
The number of stars expected within region IV is a factor 2 greater than observed. If confirmed, this discrepancy would suggest that a significant production of dust in $M \geq 4 M_{\odot}$ stars is limited to the phases close to the peak of HBB, with the highest rate of mass loss. This would demand a partial revision of the description of the massive AGB winds, which currently predicts considerables silicates formation since the early AGB phases \citep[see Fig. 5 in][]{flavia15}. 
Alternatively, we would consider possible effects of assuming the upper/lower limits of the SFH given
by Harris \& Zaritsky (2009) . The relative contribution of stars formed in different epochs assuming the standard and lower limit are almost undistinguishable, while in the upper limit case a significantly higher number of massive AGB 
stars, formed 50-100 Myr ago, would be expected. This would further exacerbate the discrepancy among the observed 
and expected number of AGB stars in region IV of the CCD, pointing in favor of the standard or the lower limit. This cannot be a definitive conclusion though, as the
low numbers involved (10 stars observed vs ~20 expected) would require a deeper analysis of the 
statistical completeness of the sources detected in region IV.
We leave this problem open.

In the CMD plane, the 
brightest oxygen--rich stars populate the region with 6 mag $< [8.0] <$ 8 mag and colours 
$1 < [3.6]-[8.0] < 2$; their position partially overlaps with the AGB stars belonging to
the sequence of C--rich objects.
The colours and magnitudes of these objects are in nice agreement with the position of 
the most obscured, O--rich stars in the samples of spectroscopically confirmed stars in the CCD and CMD by \citet{ruffle15}, \citet{anibal09} and \citet{smith95}. 
It is very interesting that the most obscured (and extreme) O-rich SMC AGB stars,
HV 1375 and IRAS 00483-7347 (spectroscopically confirmed by \citet{smith95}
and \citet{anibal09}, respectively), are inside region IV in the CCD. This seems to support our models:
i) HV 1375 is a high-luminosity, Li-rich (HBB) AGB star, which is rich in
s-process elements, like Zr and Nd, but Rb poor \citep{plez93}. The position
in the CCD and CMD diagrams (Figs. ~\ref{fsmc_ccd}-\ref{fsmc_cmd}) is consistent with our predictions for a
4-5.5 solar mass AGB experiencing HBB, something
consistent with the lack of Rb in this star (see Garc\'{\i}a--Hern\'andez et al. 2006, 2009); ii) IRAS 00483-7347 is the most massive HBB AGB star known to date
in the SMC \citep[as indicated by its extremely large Rb enhancement][]{anibal09}. Its position in the CMD, particularly the brightenss in the 8.0$\mu m$ band ($[8.0] \sim 4.7$ mag), indicates that this source is an AGB star of mass $\sim 6-7 M_{\odot}$.

We note in Fig. \ref{fsmc_ccd} the presence of two stars in the \citet{ruffle15} sample, RAW 631 and HV 12956, out of the region IV, in a zone of the CCD not covered by the evolutionary tracks (see Fig. \ref{ftracceccd}). Both stars are heavily embedded and are sufficiently red ($[8.0]-[24] > 2.39$ mag) to be classified as Far--IR objects, although they are not background galaxies, YSOs or PNe (see section 3.1.6 in Boyer et al. 2011 and section 5.1.3 in Ruffle et al. 2015). Their spectra also show evidence of crystalline silicates. 
RAW 631 present a dual chemistry, with absorption features from $C_2H_2$ and HCN and strong features from crystalline silicates, reason why it was classified as oxygen-rich by \citet{ruffle15}. These two objects present several peculiarities which renders difficult the modeling of their colors with the method we used for the other AGB stars.

Stars NGC 371, in the samples by \citet{smith95}, and HV 12149, from \citet{smith95} and \citet{ruffle15}, populating region II, were classified as oxygen--rich. We interpret these objects as descendants of stars of mass $\sim4-5.5 M_{\odot}$ that have just started to experience HBB, in the phase when the evolutionary tracks move towards region IV of the CCD (see Fig. \ref{ftracceccd})

The identification of stars in region IV of the CCD with AGB stars experiencing HBB can be
tested through spectroscopic analysis: their surface chemical composition would show the
signature of HBB, with a $C/O$ ratio below 0.05 and $^{12}C/^{13}C$ close to the
equilibrium value, $\sim 3.3$. Both these quantities can provide important information
on the strength of HBB experienced by massive AGB stars of metallicity $Z=8\times 10^{-3}$
\citep{ventura15}.

For what concerns the dust in their surroundings,
stars in this region of the CCD are surrounded by alumina dust particles of size
$\sim 0.07\mu$m and silicate grains of dimension $\sim 0.1 \mu$m; the latter particles
give the dominant contribution to the degree of obscuration of these stars, with 
optical depths in the range $0.2 < \tau_{10} < 0.8 $.

\begin{figure*}
\begin{minipage}{0.49\textwidth}
\resizebox{1.\hsize}{!}{\includegraphics{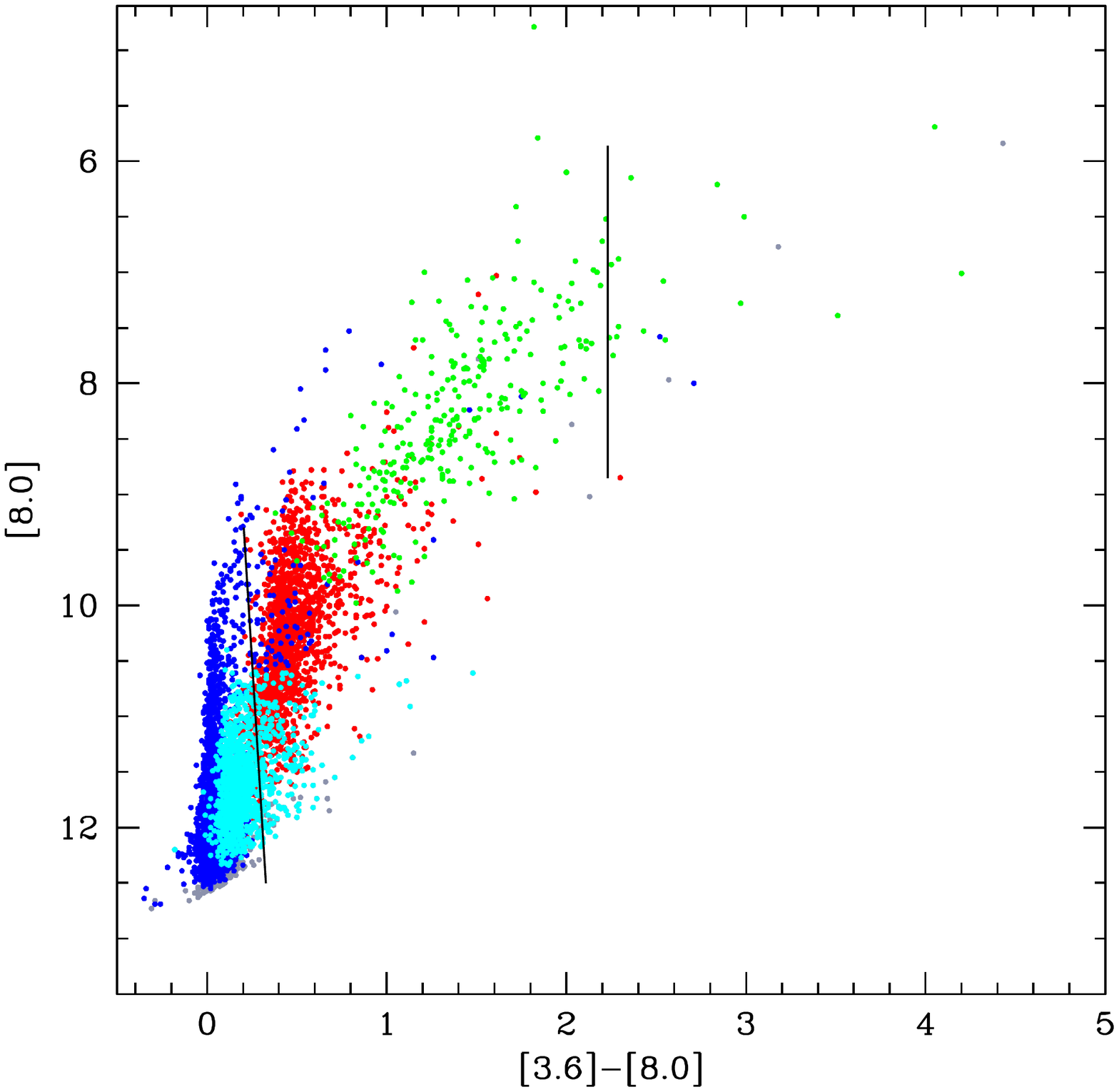}}
\end{minipage}
\begin{minipage}{0.49\textwidth}
\resizebox{1.\hsize}{!}{\includegraphics{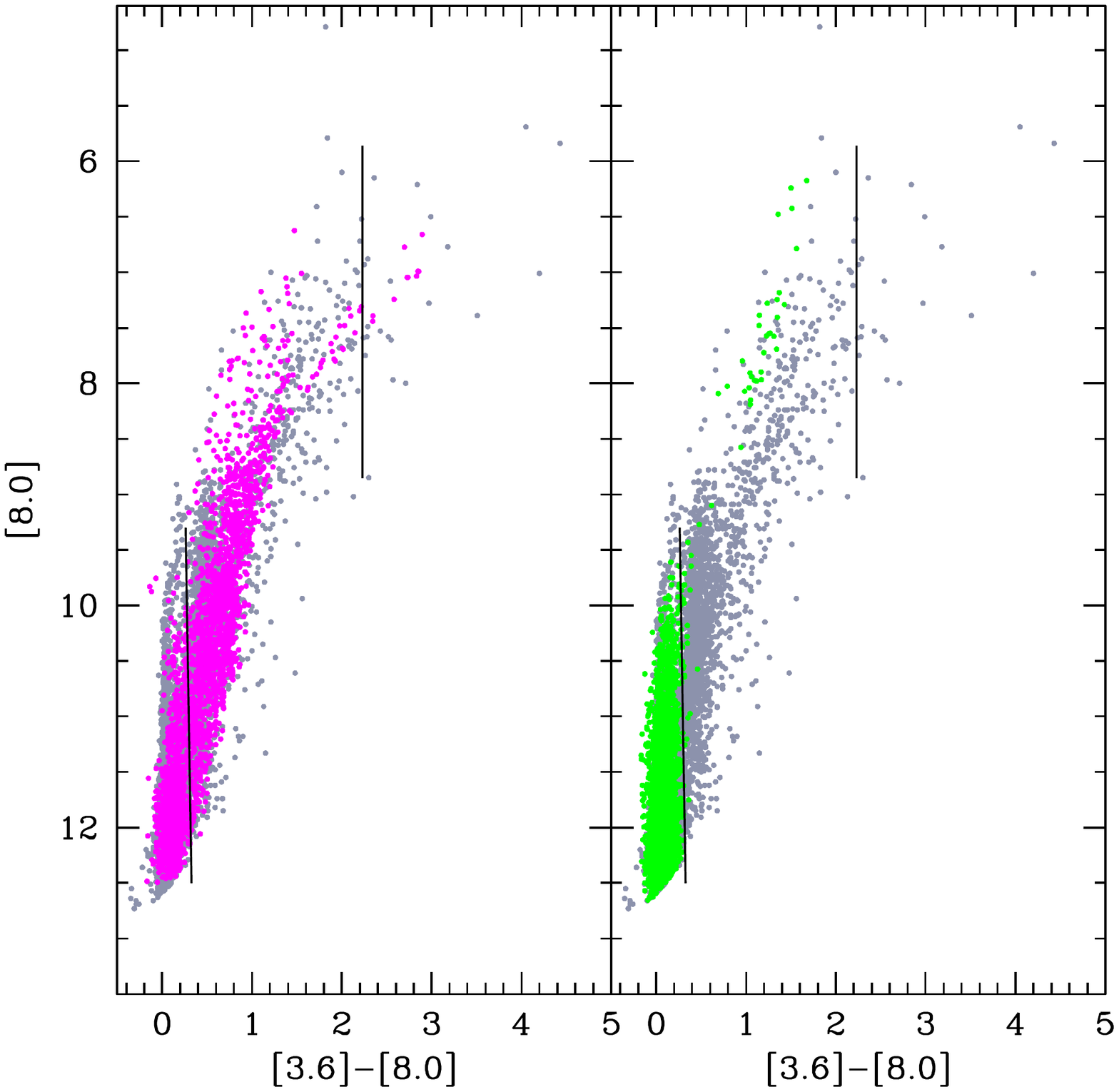}}
\end{minipage}
\vskip-50pt
\caption{Left: the distribution on the colour--magnitude ($[3.6]-[8.0], [8.0]$) plane of 
the AGB sample of the SMC by \citet{boyer11}, divided among oxygen--rich stars
(blue points), aO-AGB (cyan), carbon stars (red) and X-AGB (green). Right: the results
from our synthetic modelling in the same plane, overimposed to the observations by
\citet{boyer11}, shown with light, grey points. The simulated AGB population is 
divided among oxygen--rich (right side, green points) and carbon--rich AGB stars 
(left side, magenta).
}
\label{fboyer}
\end{figure*}

\section{A comparison with the analysis by Boyer et al. (2011)}
\citet{boyer11} divided the sample of SMC stars according to the schematisation
described in section \ref{obs}. The AGB stars in their sample were classified as  
C--rich, O--rich, X--AGB and ``aO--AGB" stars. The authors defined the latter group 
based on the position in the $J-[8]$ plane \citep[see Sec. 3.1.5 in ][]{boyer11}, where they can be distinguished  
from C--rich and O--rich AGB stars.

In the comparison among the number of stars in the various classes, the largest group of
objects, $\sim 43\%$ of the overall population, fall in the oxygen--rich sample; an 
additional $21.5\%$ are classified as aO--AGB stars. The remaining AGB stars are
divided among carbon stars ($\sim 30\%$) and X--AGB ($6\%$).

In the left panel of Fig. \ref{fboyer} we show the observations by \citet{boyer11},
in the colour--magnitude $[3.6]-[8.0], [8.0]$ plane, divided among the afore mentioned 
groups. In the right panel we show the results from
our modelling, overimposed to the observed locii: the objects which we classify as
carbon stars are shown on the left, whereas oxygen-rich stars are shown on the right.

First, we find that almost the entire X--AGB sample is composed of carbon stars; only 
a few objects classified as X--AGB, populating the brighter region of the CMD, can be
interpreted as the progeny of massive AGB stars, experiencing HBB.

In the comparison among the results from our simulations, shown in the right side of
Fig. \ref{fboyer}, and the classification introduced by \citet{boyer11} (left panel), we see
that part of the stars which we interpret as carbon stars overlap with
the aO--AGB sample and, in minor extent, with the O--rich group. 

According to our analysis the aO--AGB group introduced by \citet{boyer11} is composed of low mass stars approaching or at the beginning of the C-rich stage. Note that in a very recent work, \citet{boyer15}, on the basis of optical spectra analysis, 
claim that 50\% of the aO--AGB stars in the SMC are C--rich stars, in agreement with our interpretation. 
This leads to a slight
difference in the relative distribution of oxygen--rich AGB and carbon stars. 
According to  \citet{boyer11}, carbon stars accounts for
$\sim 36\%$ of the overall population (based on the previos point, here we added the 
carbon stars and the X--AGB), whereas according to our interpretation C--stars 
are $\sim 46\%$ of the AGB population of the SMC in the sample by \citet{boyer11}.

\begin{figure}
\resizebox{1.\hsize}{!}{\includegraphics{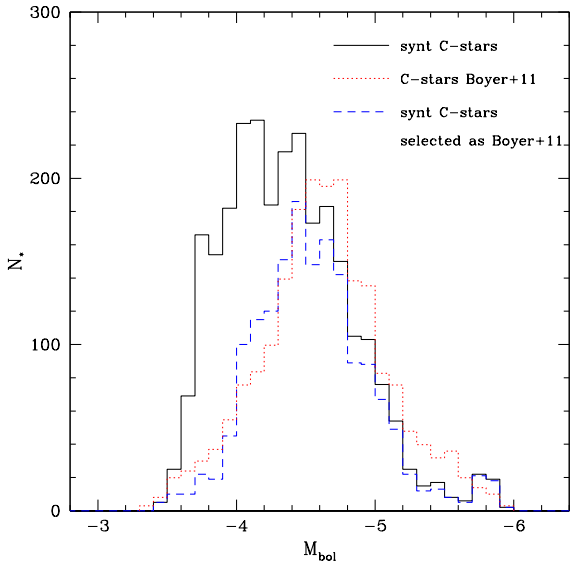}}
\vskip-10pt
\caption{The luminosity function of C--stars in the SMC found in the present work
(black solid track) and in the investigation by \citet{boyer11} (red dotted). The blue dashed line
is the luminosity function obtained with our models, when the same criterion used
by \citet{boyer11} to separate C--stars from oxygen--rich objects is adopted.
}
\label{fFL}
\end{figure}

This difference is the reason for the different luminosity function (LF) of carbon stars
between the present study and the results from \citet{boyer11}. Fig.~\ref{fFL} shows
that we predict a higher number of faint C--stars in comparison with \citet{boyer11}.
When the same criterion to separate C--stars from oxygen--rich AGB stars used by 
\citet{boyer11} is adopted, the two results become very similar. Note that on the basis of the new results 
by \citet{boyer15} quoted above, these differences would be significantly reduced.

\section{The AGB populations in the LMC and SMC}
The SAGE--LMC and SAGE--SMC surveys allowed a complete census of the AGB stars in the LMC
and SMC. Hundreds of thousands of photometric data were used to produce colour--colour
and colour--magnitude diagrams, that can be compared with results from theoretical modelling,
with the aim to characterize the sources observed.

The comparison among the infrared colours and magnitudes of the AGB stars in the LMC and SMC
shows some similarities but also significant differences, such as the
number of AGB stars in the two samples and the distribution 
of the stars observed among the various groups of the classifications introduced.

D15 focused on the AGB stars in the LMC with the largest infrared emission, called ``extreme"
\citep{riebel10, riebel12};
these stars were shown to give the largest contribution to the dust production rate of
the LMC. 
 One of the main findings of D15 was that this sample of extreme AGB stars was mainly
($\sim 95\%$) composed of C--rich stars, whereas oxygen--rich AGB give a $\sim 5\%$ contribution. This interpretation was in agreement with the conclusions by \citet{riebel12}, based on GRAMS models.
C--stars in the extreme sample (the OCS and EOCS classes introduced in D15) were identified as 
low--mass stars, whose surface chemistry has been significantly increased in carbon;
the oxygen--rich, extreme stars were interpreted as stars experiencing HBB (HBBS in D15).

In the analysis of the X-AGB population of the SMC, presented in section 5,
we also find evidence of carbon--enriched objects and of stars experiencing HBB. However,
the comparison with the LMC outlines two important differences.

First, by confronting Fig.~\ref{fsmc_ccd} with Fig.~11 in D15, we note that the sequence 
of carbon stars in the CCD, discussed in section \ref{cstars}, is much shorter in the SMC: 
while in the
LMC it extends to $[3.6]-[4.5] \sim 3$, here we have $[3.6]-[4.5] < 1.5$. The reason is
the different progenitors of the C--rich AGB stars evolving to the reddest infrared colours in the
two galaxies. D15 proposed that the EOCS sample, i.e. the carbon stars in the LMC with the
largest infrared emission, descend from stars of metallicity
$Z=8\times 10^{-3}$ and initial mass $M\sim 2.5-3~M_{\odot}$. These stars formed during the
burst in the SFR of the LMC, which occurred $\sim 500$Myr ago \citep{harris09}. The SFR
of the SMC (see Fig.~\ref{fsfh}) presents a much narrower peak at those ages, with 
practically no possibility to find stars within this range of mass in such an advanced
and short evolutionary phase; consequently,
the population of the most obscured C--stars in the SMC descend from stars of initial mass 
$\sim 1.5-1.7~M_{\odot}$ and metallicity $Z=4\times 10^{-3}$, formed $1-1.5$ Gyr ago, 
during the much longer, secondary peak in the SFR. The comparison
between the evolutionary tracks of the $1.5~M_{\odot}$ model of $Z=4\times 10^{-3}$ (middle 
panel of Fig.~\ref{ftracceccd}) and the $2.5~M_{\odot}$ model of $Z=8\times 10^{-3}$ (right 
panel of Fig.~\ref{ftracceccd}) shows that the latter evolves to much redder colours,
which is the reason for the difference in the extension of the C--star sequence 
in the observed CCD of the LMC and SMC. On the evolutionary side, as discussed in
section \ref{irspectra}, this difference is entirely due to the larger amount
of carbon accumulated at the surface of the $2.5~M_{\odot}$ model, as a consequence of
the higher number of TDUs experienced. The values reached by the optical depth in the
final AGB phases of these models is a further evidence of this behaviour: we have $\tau_{10} \sim 3$
for $(M,Z)=(2.5~M_{\odot},8\times 10^{-3})$ and $\tau_{10} \sim 0.8$
for $(M,Z)=(1.5~M_{\odot},4\times 10^{-3})$ (see middle and right panels of 
Fig.~\ref{ftau}).

Turning to the oxygen--rich AGB stars, the sample of obscured stars in the SMC, discussed in
section \ref{hbbagb}, includes a much 
smaller number of objects in comparison to the LMC. In the present work we find that
the whole sample of X--AGB is almost entirely composed of carbon stars, with a very modest
contribution from O--rich sources: Fig.~\ref{fsmc_ccd} shows that the number of stars 
populating region IV in the CCD,
where O--rich AGB stars evolve during the HBB phase (see section \ref{hbbagb}), 
is much smaller than in regions II and III, populated by
carbon stars (see section \ref{cstars}). The comparison of the SFH of the two 
galaxies provides an explanation for this dissimilarity. According to D15, the stars belonging
to the HBBS group of the LMC formed $\sim 100$Myr ago, during a peak in the SFR of the
LMC \citep{harris09}. Looking at Fig.~\ref{fsfh}, we note that the shape of the SFR of the 
SMC during the period 50--200 Myr ago is completely different: at odds with the LMC, the
SFR of the SMC presents a minimum in that period: a much smaller number of stars formed in 
that epoch, which motivates the paucity of AGB stars in region IV of the CCD.

The arguments presented here outline that the LMC is a much more favourable environment
to investigate dust production by stars evolving through the AGB phase. The LMC, compared to the SMC, harbour a higher percentage of 
dust--enshrouded AGB stars with a large infrared emission; this holds for  
oxygen--rich objects and carbon stars. In the first case, the difference among the
two galaxies is in the number of stars detected, whereas for dusty C--rich AGB stars we also
find a qualitative dissimilarity, in the infrared colours of the most obscured sources,
which are significantly redder in the LMC. 

The peculiar evolution of the SFR of the LMC is the reason for the presence of such a large percentage
of highly obscured AGB stars in this galaxy. This is because the SFR peaks in the two epochs 
($\sim 100$ Myr and $\sim 500$ Myr ago) when, based on our models of dusty AGB stars, the objects 
with the highest degree of obscuration formed. 

100 Myr is the evolution time of $\sim 5-6~M_{\odot}$ stars: this is the range of mass of stars experiencing strong HBB (temperature at bottom of the convective envelope $T > 90$ MK) during the AGB phase, with
the formation of large amounts of dust and a high infrared emission. Within the sample
of oxygen--rich stars, these are the sources which reach the largest degree of
obscuration. Models of higher mass would also evolve to extremely red colours; however,
a very few stars in this range of mass are expected, for reasons associated to the
mass function and the short duration of the AGB phase of stars in this mass range.

An earlier episode of strong star formation took place in the LMC $\sim 500$ Myr ago.
This is the formation epoch of $Z=8\times 10^{-3}$ stars of mass in the range
$2.5-3~M_{\odot}$. Among all the stars evolving through the C--star phase, these are
the stars experiencing the highest number of TDU episodes, thus accumulating the
largest quantities of carbon in the surface regions: the winds of these stars are 
therefore an extremely favourable environment for the formation of carbonaceous solid
particles, which explaines their extremely red infrared colours. Fig.~\ref{ftracceccd}
shows that the evolutionary tracks of stars in this range of mass, and metallicity 
$Z=4,8\times 10^{-3}$, are those reaching the reddest infrared colours during the
final phases of the evolution as carbon stars.

\section{Conclusions}
We study the population of AGB stars of the SMC. Our analysis
is based on the comparison between {\it Spitzer} observations and theoretical modelling
of the AGB phase. The description of the AGB evolution relies on a full integration of
the equations of stellar structure and on the description of the dust formation process 
in the circumstellar envelope; accounting for the presence of dust is crucial to
interpret the {\it Spitzer} colours of the most obscured AGB stars, the sources with the largest infrared 
emission.

In the colour--colour ($[3.6]-[4.5]$, $[5.8]-[8.0]$) diagram and in the
colour--magnitude ($[3.6]-[8.0]$, $[8.0]$) plane, we distinguish a 
population of stars with negligible infrared emission, composed mainly of 
oxygen--rich, low--mass stars of various metallicity, in the AGB phases previous to the 
C--star phase; this region of the CCD is also populated by some carbon stars with a small 
degree of obscuration, either in the early phases after the achievement of the 
C--star stage, or in the phases following each thermal pulse, before hydrogen
burning in the shell is fully restored.

The sequences of dusty AGB stars with infrared emission bifurcate in the colour--colour plane.
Carbon stars trace a diagonal band, the reddest objects having $[3.6]-[4.5] \sim 1.5$, 
$[5.8]-[8.0] \sim 1$. We interpret this population of C--rich objects as an obscuration 
sequence: the reddest sources correspond to the stars that experienced the largest number 
of Third Dredge--Up episodes, with a higher content of carbon in the surface zones. The 
spectral energy distribution of these AGB stars exhibits a large infrared emission, owing to 
the effects of the large quantities of carbonaceous dust, formed in their winds.
The carbon star sample in the SMC descend from 
stars of mass $1.5-2~M_{\odot}$ and metallicity $Z=4\times 10^{-3}$, formed between
700 Myr and 2 Gyr ago, and from a few Gyr old population of lower metallicity objects,
of mass below $1.5~M_{\odot}$. The C--rich objects of the SMC with the largest
degree of obscuration descend from stars of initial mass $\sim 1.5-1.7~M_{\odot}$ and 
metallicity $Z=4\times 10^{-3}$, formed $\sim 3$ Gyr ago, during the secondary peak in 
the SFR of the SMC.

The oxygen--rich AGB stars, surrounded by dust, populate regions in the 
colour--colour plane uncovered by the carbon rich sample. This is due to the prominent
silicate feature at $9.7\mu$m, in the spectral energy distribution, which provokes
a higher flux in the $[8.0]$ flux, for a given degree of obscuration. The oxygen--rich
stars with the largest infrared emission are interpreted as the progeny of massive
AGB stars of initial mass $\sim 5-6~M_{\odot}$ and metallicity $Z=8\times 10^{-3}$, 
experiencing HBB; these stars formed $\sim 100$ Myr ago.

The comparison among the sample of obscured AGB stars in the SMC and LMC, called ``extreme",
outlines two important differences: a) while oxygen--rich AGB stars in the LMC account for 
$\sim 5\%$ of the total sample, their contribution in the SMC is much smaller;
b) the observed sequence of the obscured carbon stars in the CCD of the LMC extends to
$[3.6]-[4.5] \sim 3$, whereas the population of C--rich objects in the SMC is
included in the region $[3.6]-[4.5] < 1.5$. 

Both these evidences indicate that dust production by AGB stars is much higher in the LMC than 
in the SMC. This is motivated by the different SFH of the two galaxies. The evolution with 
time of the SFR of the LMC proves extremely favourable to dust production by AGB stars. This originates
from the two peaks at $\sim 100$ Myr and $\sim 500$ Myr. These are the evolution times
of the sources producing the larger quantities of dust, namely the stars of mass
$\sim 5-6~M_{\odot}$, producing considerable quantities of silicates, and those with
mass $\sim 2.5-3~M_{\odot}$, which provide the largest contribution to production of
carbonaceous particles.

\section*{Acknowledgments}
The authors are indebted to the anonymous referee for the careful reading of the manuscript and for the detailed and relevant comments, that helped to increase the quality of this work. F.D. thanks F. Kemper for making available the spectroscopically confirmed sample and Martha Boyer for useful indications concerning photometrical uncertainties.
P.V. was supported by PRIN MIUR 2011 ``The Chemical and Dynamical Evolution of the Milk 
Way and Local Group Galaxies" (PI: F. Matteucci), prot. 2010LY5N2T.
D.A.G.H. acknowledges support provided by the Spanish
Ministry of Economy and Competitiveness under grant AYA2014-58082-P.
R.S. acknowledges funding from the European Research Council under the European Unions 
Seventh Framework Programme (FP/2007- 2013)/ERC Grant Agreement n. 306476.
M.D.C. acknowledges support from the Observatory of Rome.


\begin{thebibliography}{99}
\bibitem[\protect\citeauthoryear{Aringer et al.}{2009}]{grams}	
Aringer B., Girardi L., Nowotny W., Marigo P., Lederer M.~T.,
2009, A\&A, 503, 913
\bibitem[\protect\citeauthoryear{Bertoldi et al.}{2003}]{bertoldi03}
Bertoldi F. et al., 2003, A\&A, 409, L47
\bibitem[\protect\citeauthoryear{Bianchi \& Schneider}{2007}]{bianchi07}
Bianchi S, Schneider R., 2007, MNRAS, 378, 973
\bibitem[\protect\citeauthoryear{Bl\"ocker}{1995}]{blocker95}
Bl\"ocker T., 1995, A\&A, 297, 727
\bibitem[\protect\citeauthoryear{Bl\"ocker \& Sch\"oenberner}{1991}]{blocker91}
Bl\"ocker T., Sch\"oenberner D., 1991, A\&A, 244, L43
\bibitem[\protect\citeauthoryear{Bolatto}{2007}]{bolatto07}
Bolatto, A. D., Simon, J. D., Stanimirovic?, S., et al. 2007, ApJ, 655, 212
\bibitem[\protect\citeauthoryear{Bowen}{1988}]{bowen88}Bowen G.~H., 1988, ApJ, 329, 299
\bibitem[\protect\citeauthoryear{Boyer et al.}{2011}]{boyer11}
Boyer M.~L., et al., 2011, AJ, 142, 103
\bibitem[\protect\citeauthoryear{Boyer et al.}{2015}]{boyer15}
Boyer M.~L., et al., 2015, ApJ, 2015arXiv150707003B
\bibitem[\protect\citeauthoryear{Calura et al.}{2008}]{calura08}
Calura F., Pipino A., Matteucci F., 2008, A\&A, 479, 669
\bibitem[\protect\citeauthoryear{Cameron \& Fowler}{1971}]{cameron71}
Cameron A.~G.~W., Fowler W.~A., 1971, ApJ, 164, 111
\bibitem[Canuto(1992)]{canuto92} Canuto, V.~M.\ 1992, ApJ, 392, 218 
\bibitem[Canuto(1993)]{canuto93} Canuto, V.~M.\ 1993, ApJ, 416, 331
\bibitem[\protect\citeauthoryear{Canuto \& Mazzitelli}{1991}]{cm91}
Canuto V.~M.~C., Mazzitelli I., 1991, ApJ, 370, 295
\bibitem[\protect\citeauthoryear{Cioni et al.}{2000}]{cioni00}
Cioni M.-R. L., van der Marel R.~P., Loup C., Habing H. J., 2000, A\&A, 359, 601
\bibitem[\protect\citeauthoryear{Cioni et al.}{2006a}]{cioni06}
Cioni M.~R.~L., Girardi L., Marigo P., Habing H.~J., 2006, A\&A, 448, 77
\bibitem[\protect\citeauthoryear{Cioni et al.}{2006b}]{cioni06b}
Cioni M.~R.~L., Girardi L., Marigo P., Habing H.~J., 2006, A\&A, 452, 195
\bibitem[\protect\citeauthoryear{Cloutmann \& Eoll}{1976}]{cloutmann}
Cloutmann, L., \& Eoll, J.G.~1976, ApJ, 206, 548
\bibitem[\protect\citeauthoryear{De Bennassuti et al.}{2014}]{matteo14}
De Bennassuti M., Schneider R., Valiante R., Salvadori S., 
2014, MNRAS, submitted
\bibitem[\protect\citeauthoryear{Dell'Agli et al.}{2014a}]{flavia14}
Dell'Agli F., Ventura P., Garc{\'{\i}}a-Hern{\'a}ndez D.~A., Schneider R., Di Criscienzo M., 
Brocato E., D'Antona F., Rossi C., 2014a, MNRAS, 442, L38
\bibitem[Dell'Agli et al.(2014b)]{flavia14b} Dell'Agli F., 
Garc{\'{\i}}a-Hern{\'a}ndez D.~A., Rossi C., et al.\ 2014b, MNRAS, 441, 
1115
\bibitem[Dell'Agli et al.(2015)]{flavia15} Dell'Agli F., 
Ventura P., Schneider R., Di Criscienzo M., Garc{\'{\i}}a-Hern{\'a}ndez D.~A., 
Rossi C., Brocato E., \ 2015, MNRAS, 447, 2992 (D15)
\bibitem[\protect\citeauthoryear{Di Criscienzo et al.}{2013}]{paperIII}
Di Criscienzo M., Dell'Agli F., Ventura P., Schneider R., Valiante R., 
La Franca F., Rossi C., Gallerani S., Maiolino, R., 2013, MNRAS, 433, 313
\bibitem[\protect\citeauthoryear{Doherty et al.}{2014}]{doherty14}
Doherty C.~L., Gil-Pons P. Lau H.~B., Lattanzio J.~C., Siess L.,
2014, MNRAS, 437, 195
\bibitem[\protect\citeauthoryear{Dwek}{1998}]{dwek98} Dwek E., 1998, ApJ, 501, 643
\bibitem[\protect\citeauthoryear{Ferrarotti \& Gail}{2001}]{fg01}
Ferrarotti A.~D., Gail H.~P., 2001, A\&A, 371, 133
\bibitem[\protect\citeauthoryear{Ferrarotti \& Gail}{2002}]{fg02}
Ferrarotti A.~D., Gail H.~P., 2002, A\&A, 382, 256
\bibitem[\protect\citeauthoryear{Ferrarotti \& Gail}{2006}]{fg06}
Ferrarotti A.~D., Gail H.~P., 2006, A\&A, 553, 576
\bibitem[\protect\citeauthoryear{Gail \& Sedlmayr}{1985}]{gs85}
Gail H.~P., Sedlmayr E., 1985, A\&A, 148, 183
\bibitem[\protect\citeauthoryear{Gail \& Sedlmayr}{1999}]{gs99}
Gail H.~P., Sedlmayr E., 1999, A\&A, 347, 594
\bibitem[\protect\citeauthoryear{Garc\'{\i}a--Hern\'andez et al.}{2006}]{anibal06}
Garc\'{\i}a--Hern\'andez D.~A., Garc\'{\i}a-Lario P., Plez B., D'Antona F., Manchado A.,
Trigo-Rodriguez J.~M., 2006, Science, 314, 1751
\bibitem[\protect\citeauthoryear{Garc\'{\i}a--Hern\'andez et al.}{2007}]{anibal07}
Garc\'{\i}a--Hern\'andez D.~A., Garc\'{\i}a-Lario P., Plez B. et al., 2007, A\&A, 462, 711
\bibitem[\protect\citeauthoryear{Garc\'{\i}a--Hern\'andez et al.}{2009}]{anibal09}
D. A. Garc\'{\i}a--Hern\'andez, Manchado A., Lambert D. et al. 2009, ApJ 705, L31
\bibitem[\protect\citeauthoryear{Gehrz}{1989}]{gehrz89}
Gehrz R.~D., in IAU Symp. 135, Interstellar dust, ed. L. J. Allamandola \&
A. G. G. M. Tielens (Dordrecht: Kluwer), 445
\bibitem[\protect\citeauthoryear{Gordon et al.}{2011}]{gordon11}
Gordon K.~D. et al. 2011, AJ, 142, 102
\bibitem[\protect\citeauthoryear{Grevesse \& Sauval}{1998}]{gs98}
Grevesse N., Sauval A.~J, 1998, SSrv, 85, 161
\bibitem[\protect\citeauthoryear{Hanner}{1988}]{hanner88}
Hanner M., 1988, Technical report, Grain optical properties
\bibitem[Harris \& Zaritsky(2004)]{harris04} Harris J., Zaritsky D.\ 2004, AJ, 127, 1531 
\bibitem[\protect\citeauthoryear{Harris \& Zaritsky}{2009}]{harris09}
Harris J. \& Zaritsky D. 2009, ApJ, 138, 1243
\bibitem[\protect\citeauthoryear{Hauschildt et al.}{1999}]{nextgen} 
Hauschildt P.~H., Allard F., Ferguson J., Baron E., 
\& Alexander D.~R.\ 1999, ApJ, 525, 871
\bibitem[\protect\citeauthoryear{Ita et al.}{2010}]{ita10} 
Ita, Y., Onaka, T., Tanabe?, T., et al. 2010, PASJ, 62, 273
\bibitem[\protect\citeauthoryear{Keller \& Wood}{2006}]{keller06} 
Keller S.~C., Woods P.~R., 2006, ApJ, 642, 834
\bibitem[Koike et al.(1995)]{koike95} Koike, C., Kaito, C., 
Yamamoto, T., et al.\ 1995, Icarus, 114, 203
\bibitem[\protect\citeauthoryear{Maiolino et al.}{2004}]{maiolino04} 
Maiolino R., Schneider R., Oliva E., et al. 2004, Nature, 431, 533
\bibitem[\protect\citeauthoryear{Marigo}{2002}]{marigo02}
Marigo P., 2002, A\&A, 387, 507
\bibitem[\protect\citeauthoryear{Marigo \& Aringer}{2009}]{marigo09} 
Marigo P., Aringer B., 2009, A\&A, 508, 1538
\bibitem[Mazzitelli(1989)]{mazzitelli89} Mazzitelli I.\ 1989, ApJ, 340, 249
\bibitem[Mazzitelli et al. (1999)]{mazzitelli99} Mazzitelli I., D'Antona F., Ventura P., 1999, A\&A, 348, 846
\bibitem[\protect\citeauthoryear{Meixner et al.}{2006}]{meixner06} 
Meixner M. et al. 2006, AJ, 132, 2268
\bibitem[\protect\citeauthoryear{Miville-Desche�nes \& Lagache}{2005}]{miville05} 
Miville-Desche�nes, M.-A., Lagache, G. 2005, ApJS, 157, 302
\bibitem[\protect\citeauthoryear{Nanni et al.}{2013a}]{nanni13a} 
Nanni A., Bressan A., Marigo P., Girardi L., 2013a, MNRAS, 434, 488
\bibitem[\protect\citeauthoryear{Nanni et al.}{2013b}]{nanni13b} 
Nanni A., Bressan A., Marigo P., Girardi L., 2013b, MNRAS, 434, 2390
\bibitem[\protect\citeauthoryear{Nanni et al.}{2014}]{nanni14} 
Nanni A. Bressan A. Marigo P. Girardi L., 2014, MNRAS, 438, 2328
\bibitem[\protect\citeauthoryear{Nenkova et al.}{1999}]{dusty} 
Nenkova M., Ivezic Z., Elitzur M., 1999, in LPIContributions 969,
Workshop on Thermal Emission Spectroscopy and Analysis of Dust, Disks, and 
Regoliths, ed. A. Sprague, D. K. Lynch, \& M. Sitko (Houston, TX: Lunar and 
Planetary Institute), 20
\bibitem[Ordal et al.(1988)]{ordal88} 
Ordal M.~A., Bell R.~J., Alexander R.~W., Newquist L.~A., Querry M.~R. \ 1988, 
Applied Optics, 27, 1203 
\bibitem[Ossenkopf et al.(1992)]{ossenkopf92} 
Ossenkopf V., Henning T., \& Mathis J.~S.\ 1992, A\&A, 261, 567 
\bibitem[\protect\citeauthoryear{Pegourie}{1988}]{pegourie88} 
Pegourie B. 1988, A\&A, 194, 335
\bibitem[\protect\citeauthoryear{Piatti \& Geisler}{2013}]{piatti13} 
Piatti A.E., Geisler G., 2013, AJ, 145, 17
\bibitem[\protect\citeauthoryear{Pipino et al.}{2011}]{pipino11} 
Pipino A., Fan X.~L., Matteucci F., Calura F., Silva L., Granato G., Maiolino
R., 2011, A\&A, 525, A61
\bibitem[\protect\citeauthoryear{Plez et al.}{1993}]{plez93} 
Plez B., Smith V.~V.~S., Lambert D.L., 1993, ApJ, 418, 812
\bibitem[\protect\citeauthoryear{Price et al.}{2001}]{pice01} 
Price, S. D., Egan, M. P., Carey, S. J., Mizuno, D. R., Kuchar, T. A. 2001, AJ,
121, 2819
\bibitem[Riebel et al.(2010)]{riebel10} Riebel D., Meixner M., 
Fraser O., Srinivasan S., Cook K., Vijh U. \ 2010, ApJ, 723, 1195 
\bibitem[Riebel et al.(2012)]{riebel12} Riebel D., Srinivasan S., Sargent B., 
Meixner M.\ 2012, ApJ, 753, 71 
\bibitem[Ruffle et al.(2015)]{ruffle15} Ruffle P.~M.~E., Kemper F., Jones O.~C., 
et al.\ 2015, arXiv:1505.04499 
\bibitem[Sackmam \& Boothroyd (1992)]{sackmam92} Sackmam I.~J., Boothroyd A.~I.,
1992, ApJ, 392, L71 
\bibitem[\protect\citeauthoryear{Schneider et al.}{2014}]{raffa14} 
Schneider R., Valiante R., Ventura P., Dell'Agli F., Di Criscienzo M.,
Hirashita H., Kemper F., 2014, MNRAS, 442, 1440
\bibitem[\protect\citeauthoryear{Schwering \& Israel}{1989}]{schwering89} 
Schwering, P. B. W., Israel, F. P. 1989, A\&AS, 79, 79
\bibitem[\protect\citeauthoryear{Srinivasan et al.}{2011}]{srinivasan11} 
Srinivasan S., Sargent B.~A., Meixner M., 2011, A\&A, 532, A54
\bibitem[\protect\citeauthoryear{Smith et al.}{1995}]{smith95}
Smith V. V., Plez B., Lambert D., 1995, AJ, 441, 735
\bibitem[\protect\citeauthoryear{Valiante et al.}{2009}]{valiante09}
Valiante R., Schneider R., Bianchi S., Andersen A., Anja C., 2009, MNRAS, 397, 1661
\bibitem[\protect\citeauthoryear{Valiante et al.}{2011}]{valiante11}
Valiante R., Schneider R., Salvadori S., Bianchi S., 2011, MNRAS, 416, 1916
\bibitem[\protect\citeauthoryear{Ventura \& D'Antona}{2005a}]{vd05a}
Ventura P., D'Antona F., 2005a, A\&A, 431, 279
\bibitem[\protect\citeauthoryear{Ventura \& D'Antona}{2005b}]{vd05b}
Ventura P., D'Antona F., 2005b, A\&A, 439, 1075
\bibitem[\protect\citeauthoryear{Ventura \& D'Antona}{2009}]{ventura09} 
Ventura P., D'Antona F., 2009, MNRAS, 499, 835
\bibitem[\protect\citeauthoryear{Ventura, \& D'Antona}{2011}]{ventura11}
Ventura P., D'Antona F., 2011, MNRAS, 410, 2760
\bibitem[\protect\citeauthoryear{Ventura et al.}{2014a}]{paperIV} 
Ventura P., Dell'Agli F., Di Criscienzo M., Schneider R., Rossi C., La Franca F., 
Gallerani S., Valiante R., 2014a, MNRAS, 439, 977
\bibitem[\protect\citeauthoryear{Ventura et al.}{2013}]{ventura13} 
Ventura P., Di Criscienzo M., Carini R., D'Antona F., 2013, MNRAS, 431, 3642
\bibitem[Ventura et al.(2014b)]{ventura14} Ventura P., Di Criscienzo M., D'Antona F., 
Vesperini E., Tailo M., Dell'Agli F., D'Ercole A, 2014b, MNRAS, 437, 3274 
\bibitem[\protect\citeauthoryear{Ventura et al.}{2012a}]{paperI} 
Ventura P., Di Criscienzo M., Schneider R., Carini R., Valiante R., D'Antona F., 
Gallerani S., Maiolino R., Tornamb\'e A., 2012a, MNRAS, 420, 1442
\bibitem[\protect\citeauthoryear{Ventura et al.}{2012b}]{paperII} 
Ventura P., Di Criscienzo M., Schneider R., Carini R., Valiante R., D'Antona F., 
Gallerani S., Maiolino R., Tornamb\'e A., 2012b, MNRAS, 424, 2345
\bibitem[Ventura et al.(2015)]{ventura15} Ventura P., Karakas A.~I., 
Dell'Agli F., Boyer M.~L., Garc{\'{\i}}a-Hern{\'a}ndez D.~A., Di Criscienzo M.,
Schneider R., \ 2015, MNRAS, 450, 3181
\bibitem[\protect\citeauthoryear{Ventura \& Marigo}{2009}]{vm09} Ventura P.,
Marigo P., 2009, MNRAS, 399, L54
\bibitem[\protect\citeauthoryear{Ventura \& Marigo}{2010}]{vm10} Ventura P.,
Marigo P., 2010, MNRAS, 408, 2476
\bibitem[\protect\citeauthoryear{Ventura et al.}{1998}]{ventura98} Ventura P.,
Zeppieri A., Mazzitelli I., D'Antona F., 1998, A\&A, 334, 953
\bibitem[\protect\citeauthoryear{Wachter et al.}{2002}]{wachter02} 
Wachter A., Schr\"oder K.~P., Winters J.~M., Arndt T.~U., Sedlmayr E., 2002, A\&A, 384, 452
\bibitem[\protect\citeauthoryear{Wachter et al.}{2008}]{wachter08} 
Wachter A., Winters J.~M., Schr\"oder K.~P., Sedlmayr E., 2008, A\&A, 486, 497
\bibitem[\protect\citeauthoryear{Wang et al.}{2008}]{wang08} 
Wang R., Carilli C.~L., Wagg J., Bertoldi F., Walter F., Mentem K.~M., Omont A., 
Cox P., Strauss M.~A., Fan X., ert al., 2008, ApJ, 687, 848
\bibitem[\protect\citeauthoryear{Wang et al.}{2013}]{wang13} 
Wang R., Wagg J., Carilli C.~L., Walter F., Lentati L., Fan X., Riechers D.~A., 
Bertoldi F., 2013, ApJ, 773, 44
\bibitem[\protect\citeauthoryear{Westerlund}{1997}]{westerlund97} 
Westerlund B.~E., 1997, The Magellanic Clouds (Cambridge Astrophysics series, Vol.~29;
Cambridge: Cambridge Univ. Press)
\bibitem[\protect\citeauthoryear{Wilke et al.}{2003}]{wilke03} 
Wilke, K., Stickel, M., Haas, M., et al. 2003, A\&A, 401, 873
\bibitem[Woods et al.(2011)]{woods11} Woods, P.~M., Oliveira, 
J.~M., Kemper, F., et al.\ 2011, MNRAS, 411, 1597 
\end{thebibliography}
\end{document}